\pdfoutput=1
\documentclass[11pt,a4]{article}
\usepackage{array}
\usepackage{a4wide}
\usepackage[hidelinks]{hyperref}
\usepackage{latexsym}
\usepackage{cite}
\usepackage{mathrsfs}
\usepackage{amssymb}
\usepackage{amsmath}
\usepackage{graphicx}
\usepackage{braket}
\usepackage{enumerate}
\usepackage[usenames,dvipsnames]{xcolor}
\usepackage{todonotes}

\usepackage{float}
\usepackage{mathtools}

\usepackage{adjustbox}

\graphicspath{ {Images/} }

\newcommand{\N}{\mathcal{N}}
\newcommand{\M}{\mathcal{M}}
\newcommand{\J}{\mathcal{J}}

\newcommand{\dotprod}{\!\cdot\!}
\newcommand{\req}[1]{\!(\ref{#1})}

\newcommand{\anbr}[2]{\langle #1\,#2\rangle}
\newcommand{\sqbr}[2]{[#1\,#2]}
\newcommand{\ang}[1]{\left\langle #1\right\rangle}

\newcommand{\twbr}[4]{\langle#1\,#2\,#3\,#4\rangle}

\newcommand{\br}[1]{\langle#1\rangle}

\newcommand{\ltilde}[1]{\widetilde{\lambda}_{#1}}

\DeclareMathOperator*{\res}{Res}
\DeclareMathOperator{\gl}{GL}

\newcommand{\eq}[1]{\begin{equation}#1\end{equation}}
\newcommand{\eqs}[1]{\begin{equation}\begin{split}#1\end{split}\end{equation}}

\date{}
\begin{document}

\title{$\mathcal{N}=7$ On-Shell Diagrams and Supergravity Amplitudes in Momentum Twistor Space}
\author{Connor Armstrong, Joseph A. Farrow and Arthur E. Lipstein\vspace{7pt}\\ \normalsize \textit{
Department of Mathematical Sciences}\\\normalsize\textit{Durham University, Durham, DH1 3LE, United Kingdom}}
\maketitle
\begin{abstract}
We derive an on-shell diagram recursion for tree-level scattering amplitudes in $\mathcal{N}=7$ supergravity. The diagrams are evaluated in terms of Grassmannian integrals and momentum twistors, generalising previous results of Hodges in momentum twistor space to non-MHV amplitudes. In particular, we recast five and six-point NMHV amplitudes in terms of $\mathcal{N}=7$ R-invariants analogous to those of $\mathcal{N}=4$ super-Yang-Mills, which makes cancellation of spurious poles more transparent. Above 5-points, this requires defining momentum twistors with respect to different orderings of the external momenta.

\end{abstract}

\pagebreak
\tableofcontents



\section{Introduction}

In recent years, new geometric descriptions have been discovered for scattering amplitudes such as the Amplituhedron for planar $\mathcal{N}=4$ super-Yang-Mills (SYM) \cite{Arkani-Hamed:2013jha,Arkani-Hamed:2017vfh}. The earliest hints of such a description were discovered by Hodges, who showed that 6-point NMHV amplitudes in pure Yang-Mills theory can be described as polytopes in momentum twistor space, making the cancellation of spurious poles completely manifest \cite{Hodges:2009hk}. This description was extended to all NMHV amplitudes in $\mathcal{N}=4$ SYM in \cite{ArkaniHamed:2010gg}, where an amplitude with $k$ negative helicity particles is referred to as N$^{k-2}$MHV. Another important perspective came from on-shell diagrams \cite{Arkani-Hamed:2016byb}, which provide a diagrammatic representation of BCFW recursion \cite{Britto:2005fq,ArkaniHamed:2010kv,Boels:2010nw} and give rise to Grassmannian integral formulas. A crucial feature of this representation is that all tree-level amplitudes and loop integrands in planar $\mathcal{N}=4$ SYM have a so-called dlog form, which was simultaneously discovered using Wilson loops in momentum twistor space \cite{Lipstein:2012vs,Lipstein:2013xra}. On-shell diagrams for $\mathcal{N}<4$ SYM and non-planar $\mathcal{N}=4$ SYM were subsequently studied in \cite{Benincasa:2015zna,Benincasa:2016awv} and \cite{Arkani-Hamed:2014bca,Franco:2015rma,Bourjaily:2016mnp}, respectively. Ultimately, the tree-level amplitudes and planar integrands of $\mathcal{N}=4$ were identified with differential forms having logarithmic singularities on the boundaries of a new mathematical object generalising the positive Grassmannian known as the Amplituhedron. A review of the Amplituhedron and related developments can be found in \cite{Ferro:2020ygk}. 

One of the most significant implications of the Amplituhedron is that physical principles such as unitarity and locality appear to have a purely geometric origin, at least in the context of planar $\mathcal{N}=4$ SYM. Such a description would therefore be profound in the context of gravitational amplitudes, but has so far remained elusive despite some progress. In particular, decorated on-shell diagrams encoding BCFW recursion relations for tree-level $\mathcal{N}=8$ supergravity (sugra) amplitudes \cite{Bedford:2005yy,Cachazo:2005ca} were developed in \cite{Heslop:2016plj}. As in $\mathcal{N}=4$ SYM, these objects give rise to Grassmannian integral formulas, although they do not generically have a dlog form and the extension to loop-level amplitudes is not known, although undecorated on-shell diagrams were used to compute cuts of loop diagrams in \cite{Herrmann:2016qea}.  In seminal work by Hodges \cite{Hodges:2011wm}, a BCFW recursion relation was developed for $\mathcal{N}=7$ sugra and used to obtain momentum twistor formulas for MHV amplitudes and spinor expressions for the 6-point NMHV amplitude. This recursion relation was then used to prove a determinant formula for gravitational MHV amplitudes\cite{Hodges:2012ym}, greatly simplifying the previously known BGK formula \cite{Berends:1988zp}. The generalisation of this formula to non-MHV amplitudes was obtained using twistor string theory \cite{Skinner:2013xp,Geyer:2014fka}. The relation between twistor string and on-shell diagram diescriptions of $\mathcal{N}=8$ sugra amplitudes was explored in \cite{Farrow:2017eol}, where a Grassmannian integral formula for the 6-point NMHV amplitude was obtained for the first time.

In this paper, we will build on these developments with the goal of finding hints of geometric structure in $\mathcal{N}=7$ sugra amplitudes analogous to that of $\mathcal{N}=4$ SYM. First we recast the $\mathcal{N}=7$ recursion found by Hodges in terms of on-shell diagrams. This is similar to the $\mathcal{N}=8$ recursion developed in \cite{Heslop:2016plj} in that the on-shell diagrams are decorated and give rise to Grassmannian integral formula, but exhibits a few important differences. First of all, the $\mathcal{N}=7$ diagrams are decorated with arrows indicating helicity flow. Secondly, for a certain choice of BCFW bridge the $\mathcal{N}=7$ recursion requires fewer diagrams. For example, for MHV amplitudes there are $(n-3)!$ diagrams rather than $(n-2)!$, so the $\mathcal{N}=7$ recursion incorporates a property of sugra amplitudes known as the bonus relations \cite{ArkaniHamed:2008gz} (this property of $\mathcal{N}=7$ recursion was first observed in \cite{Liu:2014mua}). The price to pay for having fewer diagrams is that they generally contain more closed cycles which can become tedious to evaluate at high multiplicity using conventional methods, so we develop a new technique which avoids summing over closed cycles. This technique can also be applied to other theories.

We then convert these formulas to momentum twistor space, reproducing Hodges' results for MHV amplitudes and obtaining new momentum twistor formulas for non-MHV amplitudes, written in terms of $\mathcal{N}=7$ R-invariants analogous to those of $\mathcal{N}=4$ SYM. Beyond 5-points, this requires describing different subset of on-shell diagrams using momentum twistors defined with respect to different permutations of external momenta, analogous to defining local coordinates on a manifold \footnote{This possibility was first suggested to us in 2017 by Andrew Hodges.}. In addition to producing R-invariants, this way of defining momentum twistors leads to concise expressions for the spurious poles of the 6-point NMHV superamplitude, making their cancellation completely transparent. This strongly suggests the existence of a geometric explanation for the cancellation of spurious poles analogous to the one found in Yang-Mills amplitudes.  

This paper is organised as follows. In section \ref{osdsec}, we review spinor notation for superamplitudes and describe the on-shell diagram recursion for $\mathcal{N}=7$ sugra and an algorithm for evaluating $\mathcal{N}=7$ on-shell diagrams in terms of Grassmannian integrals. In section \ref{momtwist} we review the basics of momentum twistor space and describe various useful formulae for writing spinor brackets and fermionic delta functions in terms of momentum twistors. In section \ref{mhvsec} we use on-shell diagrams to compute MHV amplitudes in terms of Grassmannian integrals and momentum twistors. In section \ref{nmhvsec}, we use on-shell diagrams to compute non-MHV amplitudes, write them in terms of momentum twistor space, and analyse the cancellation of spurious poles. In section \ref{conclusion} we present our conclusions and future directions. There are also several appendices. In Appendix \ref{app:ferm}, we derive momentum twistor formulas for supermomentum delta functions. In Appendix \ref{6ptdetail} we provide additional details about the on-shell diagrams for the 6-point NMHV superamplitude. In Appendix \ref{transitionfunctions} we derive transition functions for momentum twistors defined with respect to different permutations of external momenta. Finally, in appendix~\ref{sec:alg} we describe a new computational method for simplifying expressions in momentum twistor space. We give an explicit realisation of this algorithm in {\sc Mathematica}, submitted as an auxilliary file with this publication.       

\section{On-Shell Diagrams} \label{osdsec}

First we review some standard notation for scattering amplitudes.
A null momentum in four dimensions can be expressed in the following
bi-spinor form:
\begin{equation}
p_{i}^{\alpha\dot{\alpha}}=\lambda_{i}^{\alpha}\tilde{\lambda}_{i}^{\dot{\alpha}},
\label{spinor}
\end{equation}
where $\alpha\in\left\{ 1,2\right\} $, $\dot{\alpha}\in\left\{ \dot{1},\dot{2}\right\} $
and $i$ is a particle label. In terms of spinors,
one defines the following inner products
\begin{equation}
\left\langle ij\right\rangle =\epsilon_{\alpha\beta}\lambda_{i}^{\alpha}\lambda_{j}^{\beta},\,\,\,\left[ij\right]=\epsilon_{\dot{\alpha}\dot{\beta}}\tilde{\lambda}_{i}^{\dot{\alpha}}\tilde{\lambda}_{j}^{\dot{\beta}},
\label{2brackets}
\end{equation}
where $\epsilon$ is the anti-symmetric Levi-Cevita symbol. To simplify
notation, we will omit spinor indices. A very useful identity for
spinor manipulations is the Schouten identity:
\begin{equation}
\left\langle ij\right\rangle \lambda_{k}+{\rm cyclic}=0.
\label{schouten}
\end{equation}
In supersymmetric theories, the supermomentum is defined as 
\begin{equation}
q_{i}^{\alpha a}=\lambda_{i}^{\alpha}\eta_{i}^{a},
\label{superm}
\end{equation}
where $\eta$ is a Grassmann variable and $a=1,...,\mathcal{N}$,
where $\mathcal{N}$ indicates the amount of supersymmetry, i.e. $\mathcal{N}=4,8$
for maximal SYM and sugra, respectively.
Supermomentum conservation is imposed by the delta function $\delta^{(4|2\mathcal{N})}(P|Q)$,
where $P=\sum_{i}p_{i}$ and $Q=\sum_i q_{i}$. It will be convenient to factor out the supermomentum delta function from scattering amplitudes, and we will denote the resulting quantity with a bar: 
\begin{equation}
\mathcal{M}=\delta^{(4|14)}(P|Q)\overline{\mathcal{M}},
\label{ampbar}
\end{equation}
where we have set $\mathcal{N}=7$.

It is convenient to define scattering amplitudes in supersymmetric theories in terms of superfields. In $\mathcal{N}=8$ supergravity, the superfield takes the form 
\eq{
\Phi=h^{+}+...+h^{-}\eta^{8},
}
where $h^\pm$ are the two helicity states of the graviton, and the ellipsis denote the on-shell states of lower spin bosonic and fermionic fields. Note that the superfield is expanded in the Grassmann variable $\eta$ and therefore truncates at eighth order. For $\mathcal{N}=7$, there are two supermultiplets which contain the positive and negative helicity states of the graviton, respectively:
\eq{
\Phi^{+}=\left.\Phi\right|_{\eta^{8}=0},\,\,\,\Phi^{-}=\int d\eta^{8}\Phi.
}
Note that $\mathcal{N}=7$ sugra has the same field content as $\mathcal{N}=8$ and they are perturbatively equivalent. From this, we also see that an N$^{k-2}$MHV amplitude (whose graviton component has $k$ negative helicity gravitons) has fermionic degree $7k$. Using the relation between superfields, it is straightforward to extract $\mathcal{N}=7$ superamplitudes from $\mathcal{N}=8$. For example, at three points we have
\eqs{
\mathcal{M}^{\mathcal{N}=7}_3(--+)&=\int d\eta_{1}^{8}d\eta_{2}^{8}\left.\mathcal{M}^{\mathcal{N}=8}_3\right|_{\eta_{3}^{8}=0},\\
\mathcal{M}^{\mathcal{N}=7}_3(++-)&=\int d\eta_{3}^{8}\left.\overline{\mathcal{M}}^{\mathcal{N}=8}_3\right|_{\eta_1^{8}=\eta_{2}^{8}=0}
}
where $\mathcal{M}_3$ and $\overline{\mathcal{M}}_3$ are the 3-point MHV and $\overline{\rm{MHV}}$ amplitudes of $\mathcal{N}=8$ sugra, respectively. An $n$-point $\overline{\rm{MHV}}$ amplitude describes the scattering of two positive helicity gravitons and $n-2$ negative helicity gravitons. 

Using three-point amplitudes, we can build all higher point tree-level amplitudes using BCFW recursion \cite{Britto:2005fq,Bedford:2005yy,Cachazo:2005ca}. In this approach, one deforms the supermomenta of two legs in a way that preserves momentum conservation and their on-shell properties. The deformed amplitude then develops poles in the deformation parameter, whose residues correspond to products of deformed lower-point amplitudes. This recursion was generalised to all-loop integrands in planar $\mathcal{N}=4$ SYM in \cite{ArkaniHamed:2010kv}. See \cite{Farrow:2020voh} for recent progress on loop-level recursion in generic theories. 

On-shell diagrams provide a diagrammatic representation of BCFW recursion. They are built out of black and white vertices denoting MHV and $\overline{\rm{MHV}}$ amplitudes respectively, which are connected by lines representing an integral over on-shell states. This was first developed for planar $\mathcal{N}=4$ SYM \cite{Arkani-Hamed:2016byb} and later generalised to tree-level amplitudes of $\mathcal{N}=8$ sugra \cite{Heslop:2016plj}. These can be obtained by decorating planar on-shell diagrams and summing over permutations of unshifted legs, giving $(n-2)!$ terms. It is also possible to carry out the $\mathcal{N}=8$ recursion in other ways which will generically involve non-planar diagrams. 

In this paper, we will develop on-shell diagrams for $\mathcal{N}=7$ sugra. Since there are two supermultiplets encoding the two graviton helicities, the edges of the diagrams will be labelled with arrows to indicate helicity flow (which should not be confused with momentum flow). In particular, incoming arrows denote the negative helicity supermultiplet, and outgoing arrows denote the positive helicity supermultiplet. Hence, 3-point MHV amplitudes will have two incoming arrows and one outgoing arrow, while 3-point $\overline{\rm{MHV}}$ amplitudes have two outgoing arrows and one incoming arrow. Three-point superamplitudes and edges are given by:
\begin{align}
\vcenter{\hbox{\includegraphics[width=1.9cm]{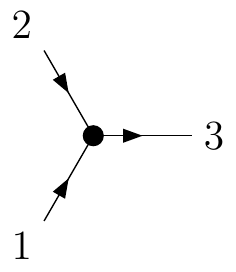}}}&\equiv \br{12} \frac{\delta^{14}(\lambda_1\eta_1 + \lambda_2\eta_2 + \lambda_3\eta_3)}{\br{12}^2\br{23}^2\br{31}^2},\\
\vcenter{\hbox{\includegraphics[width=1.9cm]{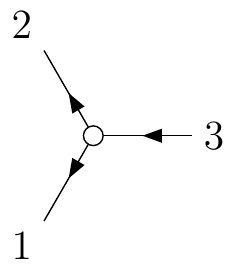}}}&\equiv [12] \frac{\delta^{7}([12]\eta_3 + [23]\eta_1 + [31]\eta_2)}{[12]^2[23]^2[31]^2},\\
\vcenter{\hbox{\includegraphics[width=1.9cm]{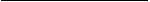}}}&\equiv \int\frac{d^2\lambda\, d^2\ltilde{}}{\gl(1)}d^7\eta.
\end{align}
Momentum conservation implies that $\tilde{\lambda}_1\propto \tilde{\lambda}_2 \propto \tilde{\lambda}_3$ for an MHV vertex and $\lambda_1\propto \lambda_2 \propto \lambda_3$ for an $\overline{\rm{MHV}}$ vertex.

All tree-level amplitudes can be constructed from 3-point vertices using the following diagrammatic recursion:
\begin{equation}
\vcenter{\hbox{\includegraphics[width=2.8cm]{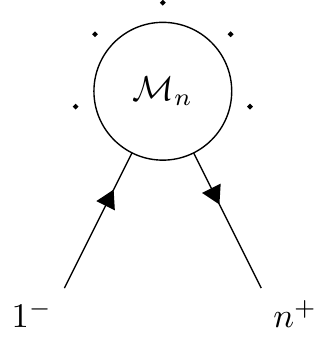}}}= \sum_{L,R}\vcenter{\hbox{\includegraphics[width=4.5cm]{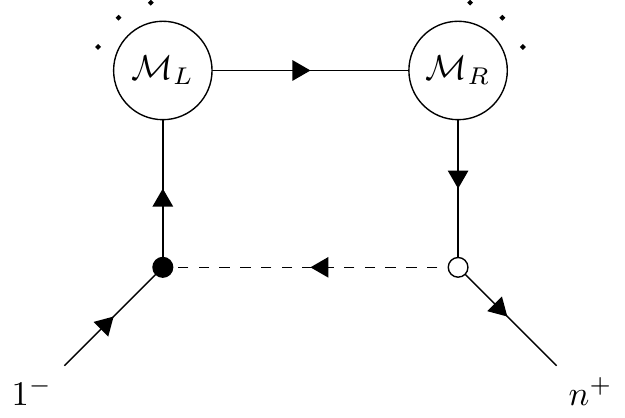}}},
\label{eqn:recursion}
\end{equation}
\begin{equation}
\mathrm{where} \vcenter{\hbox{\includegraphics[width=4.5cm]{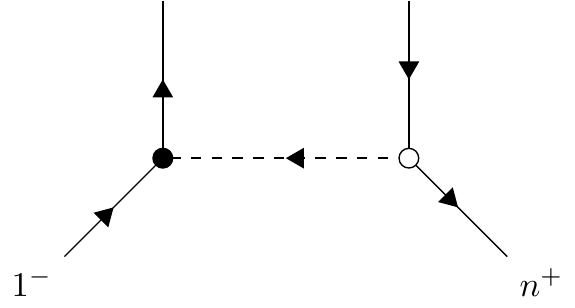}}} = \frac{1}{p_1\dotprod p_n}\vcenter{\hbox{\includegraphics[width=4.5cm]{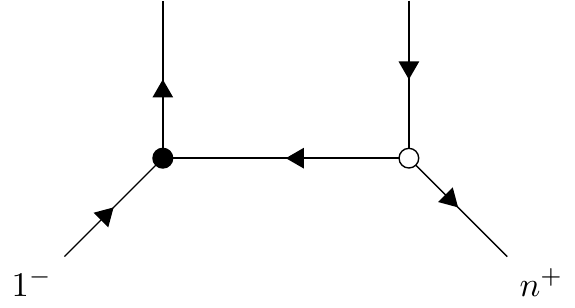}}}.
\label{eqn:decorationdefn}
\end{equation}
The object connecting the lower-point amplitudes on the right-hand-side of \eqref{eqn:recursion} implements a BCFW shift of legs 1 and $n$ and is known as a BCFW bridge. The external arrows of the bridge imply a $\langle1^-,n^+]$ shift and the internal arrows are fixed by helicity flow. Choosing this BCFW bridge at every step of the recursion always produces on-shell diagrams with closed cycles and the choice also fixes their orientation. The sum in equation~(\ref{eqn:recursion}) is over all partitions of particles $\{2,\dots,n-1\}$ into sets $L$ and $R$. The recursion is carried out such that we always feed the fixed legs of the subamplitudes into the recursion. Using the BCFW bridge in \eqref{eqn:decorationdefn}, the $\mathcal{N}=7$ recursion leads to fewer diagrams than the $\mathcal{N}=8$ recursion\cite{Heslop:2016plj}. For example, at MHV there are $(n-3)!$ instead of $(n-2)!$ terms so it appears that the $\mathcal{N}=7$ recursion encodes bonus relations \cite{ArkaniHamed:2008gz}, as was previously observed in \cite{Liu:2014mua}. In \cite{He:2010ab}, the bonus relations were used to write non-MHV amplitudes in $\mathcal{N}=8$ sugra as a sum over $(n-3)!$ terms. It would be interesting to investigate how this cmpares to the $\mathcal{N}=7$ recursion. 

Whereas $\mathcal{N}=7$ sugra amplitudes are well-behaved when the BCFW deformation is taken to infinity, this is not the case for $\mathcal{N} < 4$ SYM amplitudes \cite{Lal:2009gn,Elvang:2011fx}, so a different choice of BCFW bridge should be used in those theories. A $\langle1^-,n^+]$ shift was considered in $\mathcal{N}=3$ SYM using on-shell diagrams in \cite{Arkani-Hamed:2016byb}, where it was shown that the correct result could be obtained at 4-points by summing over orientations of closed cycles, but we find that this prescription does not work at higher points.

\begin{figure}[H]
\centering
\includegraphics[width=\textwidth]{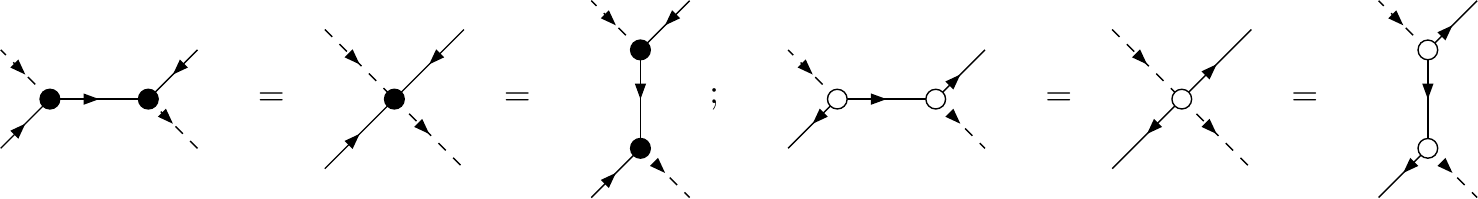}
\caption{Merging rules for $\N=7$ on-shell diagrams. The decorated edges must appear opposite.}
\label{fig:mergers}
\end{figure}

\begin{figure}[H]
\centering
\includegraphics[width=5cm]{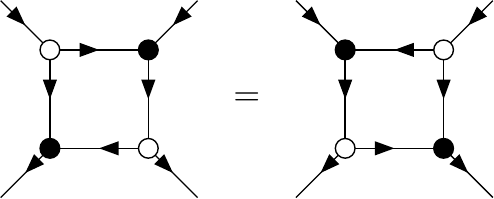}
\caption{Square move for $\N=7$ on-shell diagrams.}
\label{fig:SquareMove}
\end{figure}

$\N=7$ on-shell diagrams enjoy a number of equivalence relations similar to those of $\mathcal{N}=8$ sugra. In particular mergers appear with an orientation encoding helicity flow, as shown in Figure \ref{fig:mergers}, and square moves are only valid for diagrams where the incoming arrows are adjacent, shown in Figure \ref{fig:SquareMove}. We are also free to move decorations to the opposite edge of a box, which can be seen from the definition in \eqref{eqn:decorationdefn}.

\subsection{Evaluating on-shell diagrams}
\label{sec:algorithm}

On-shell diagrams for $n$-point N$^{k-2}$MHV amplitudes naturally give rise to integrals over the space of $k$ planes in $n$ dimensions, also known as the Grassmannian Gr$(k,n)$. These integrals are represented as an integral over a $k\times n$ matrix $C$ modulo a left action of GL$(k)$. The rows of $C$ are associated with the external legs with a $\Phi^-$ multiplet, or equivalently legs with incoming arrows, while the columns are associated will all external legs. As we describe below, the $C$ matrix for a given on-shell diagram can be computed by assigning edge variables and summing over paths through the diagram taking the product of the edge variables encountered along each path. One then lifts the integral over edge variables to a covariant contour integral in the Grassmannian. A detailed algorithm for evaluating on-shell diagrams in $\mathcal{N}=8$ sugra can be found in \cite{Heslop:2016plj} and Appendix A of \cite{Farrow:2017eol}. The algorithm for $\mathcal{N}=7$ is very similar, so we will describe it below more schematically:

\begin{enumerate}
\item Label every half edge (including external edges) with an edge variable $\alpha$. Set one of the two edge variables on each internal edge to unity and set one of the remaining edge variables associated with each vertex to unity, leaving $2n - 4$ edge variables.
\item Include $d\alpha/\alpha^2$ for each edge entering a black vertex or leaving a white vertex and $d\alpha/\alpha^3$ for each edge leaving a black vertex or entering a white vertex.
\item Multiply by $\br{ij}$ for each black vertex and $[ij]$ for each white vertex, where $i, j$ are the two incoming or outgoing legs, respectively.
\item Include a kinematic factor to each decorated edge as shown in \eqref{eqn:decorationdefn}.
\item To relate internal and external spinors, sum over paths according to
\begin{align}
\ltilde{i} &= \sum_{\mathrm{paths}\ i\to j}\left(\prod_{\mathrm{edges\ in\ path}}\alpha_e\right)\ltilde{j},\label{paths1}\\
\lambda_{i} &= \sum_{\mathrm{paths}\ j\to i}\left(\prod_{\mathrm{edges\ in\ path}}\alpha_e\right)\lambda_j.
\label{paths2}
\end{align}
The matrix element $C_{ij}$ can then be computed by summing over all paths from leg $i$ to leg $j$, taking the product of all the edge variables encountered along each path as in \eqref{paths1}. 
\item If there is a closed cycle, one will need to sum an infinite series when computing the $C$-matrix in the previous step. Moreover, one will need to include the factor $\mathcal{J}_C^{\N-4} = \mathcal{J}_C^3$, where $\J_C$ comes from a sum over products of disjoint cycles \cite{Herrmann:2016qea}:
\begin{equation}
\J_C = 1 + \sum_i f_i + \sum_{\mathrm{disjoint}\ i,j}f_i f_j + \sum_{\mathrm{disjoint}\ i,j,k}f_i f_j f_k + ...,
\end{equation}
where $f_i$ is minus the product of edge variables around the i$^{\mathrm{th}}$ cycle. In section \ref{closedcycle} we will describe an alternative method which avoids summing over closed cycles when computing the $C$-matrix, and automatically computes $\J_C$. 
\item Include $\delta^{k(2|7)}(C\cdot \ltilde{})\delta^{2\times (n-k)}(\lambda\cdot C^\perp)$ where $C^\perp$ is an $n\times (n-k)$ matrix satisfying $C \cdot C^{\perp}=0$, whose matrix elements can be computed by summing over the reverse paths in \eqref{paths2}. The dot products appearing in the delta functions are with respect to particle number, so for example $C\cdot \tilde{\lambda}$ can be written more explicitly as $\sum_{i=1}^{n}C_{Ij}\tilde{\lambda}_{j}^{\alpha}$, where $I$ labels the $k$ rows. After taking into account momentum conservation, there are $2n-4$ bosonic delta functions, which precisely matches the number of edge variables. The argument of the fermionic delta functions is $C\cdot \eta$, which we suppress for brevity. The resulting integral over edge variables can be thought of as a gauge-fixed Grassmannian integral formula, where the gauge symmetry is GL$(k)$. 
\item Covariantise the integral over edge variables to an integral over $\mathrm{Gr}(k,n)$ by writing the edge variables in terms of minors of the $C$ and $C^\perp$ matrices. Since $\mathrm{Gr}(k,n)$ has dimension $k(n-k)$ but there are only $2n-4$ edge variables, this will imply a contour in the Grassmannian of dimension $k(n-k)-(2n-4)=(k-2)(n-k-2)$.
\end{enumerate}

\subsection{Closed Cycles} \label{closedcycle}

\begin{figure}[h]
\centering
\includegraphics[width=3cm]{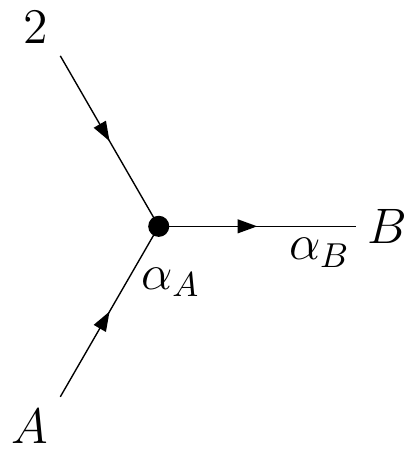}
\caption{Vertex showing example of how to truncate geometric series. Here we take leg 2 to be external (with the corresponding edge variable set to 1) and legs $A$ and $B$ to be internal.}
\label{fig:proptrick}
\end{figure}

The $\mathcal{N}=7$ recursion has fewer diagrams compared to $\mathcal{N}=8$, but the price to pay is that they generally contain more closed cycles, which can become very cumbersome to evaluate following step 5 of the algorithm in the previous subsection. This technical difficulty can be overcome as follows. Instead of expanding incoming $\ltilde{i}$ in terms of only outgoing $\ltilde{j}$, we instead expand in terms of all external $\ltilde{j}$. In particular, when summing over paths originating from each incoming leg as described in step 5, we truncate the path if we reach another incoming leg and then write the last internal spinor along the path in terms of the external spinor using the fact that all $\tilde{\lambda}$ spinors are proportional at a black vertex. For example, in Figure \ref{fig:proptrick} let us suppose that A and B are internal edges while $2$ is an external edge. When computing the $C$-matrix according to step 5 of the algorithm, we would write $\ltilde{A} = \alpha_A\ltilde{B}$ and then continue to expand $\ltilde{B}$. We instead truncate the path by writing $\ltilde{A} = \alpha_A\ltilde{2}$. The resulting $C$ matrix will have non-zero elements between incoming legs, but we can find a GL($k$) transformation to set the columns of the incoming legs to a unit matrix. We will refer to the $C$-matrix before this gauge-fixing as $\tilde{C}$. This will yield the same result as step 5 of the algorithm without having to sum over closed cycles. Moreover the Jacobian in step 6 is given by the inverse of the determinant of the GL$(k)$ transformation. This method can also be applied to on-shell diagrams in other theories.

\section{Momentum Twistors} \label{momtwist}

\begin{figure}[H]
\centering
\includegraphics[width=9cm]{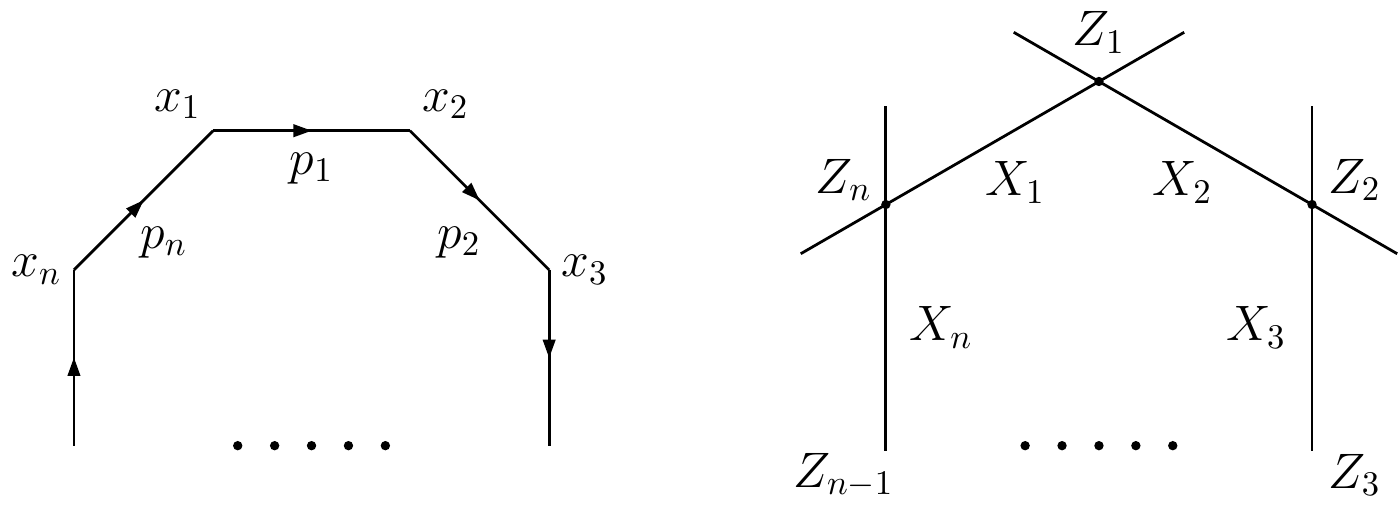}
\caption{Diagrams indicating the relation between momenta $p_i$, dual variables $x_i$ and momentum twistors $Z_i$.}
\label{fig:WilsonLoop}
\end{figure}

In this section, we review some basic properties of momentum twistors, which were first defined in \cite{Hodges:2009hk}. These variables are defined in terms of region momentum coordinates $x_i$ defined by the constraints 
\begin{equation}
\left(x_{i}-x_{i+1}\right)^{\alpha\dot{\alpha}}=p_{i}^{\alpha\dot{\alpha}}.\label{eq:region}
\end{equation}
Region momentum coordinates can be visualised as the vertices of a null polygon constructed
by arranging the external momenta head-to-tail, as depicted in Figure \ref{fig:WilsonLoop}. The conformal group in region momentum space is known as the dual conformal group, which was shown to be a symmetry of planar $\mathcal{N}=4$ SYM \cite{Drummond:2006rz,Brandhuber:2008pf,Drummond:2008vq} and a superconformal Chern-Simons theory known as the ABJM theory \cite{Bargheer:2010hn,Huang:2010qy,Gang:2010gy}. Although sugra amplitudes do not enjoy dual conformal symmetry, we will see that momentum twistors are nevertheless useful for describing them. Momentum twistors are then defined as
\eq{
Z_{i}^{A}=\left(\lambda_{i}^{\alpha},\mu_{i}^{\dot{\alpha}}\right),
}
where $A\in\left\{ 1,..,4\right\} $ are indices in the fundamental
representation of the dual conformal group $SU(4)$, $\lambda_{i}$
is the spinor associated with the momentum of particle $i$ according
to \eqref{spinor}, and $\mu_{i}$ is an auxiliary spinor satisfying the incidence
relation $\mu_{i}=x_{i}\cdot\lambda_{i}$. Using the incidence relation
and \eqref{eq:region}, one finds that
\begin{equation}
x_{i}^{\alpha\dot{\alpha}}=\frac{\lambda_{i}^{\alpha}\mu_{i-1}^{\dot{\alpha}}-\lambda_{i-1}^{\alpha}\mu_{i}^{\dot{\alpha}}}{\left\langle i-1i\right\rangle },\label{line}
\end{equation}
which implies that a point $x_{i}$ in region momentum space corresponds
to a line in momentum twistor space associated with $Z_{i-1}$ and
$Z_{i}$. Using this correspondence, a null polygon in momentum space
can be mapped into a polygon in momentum twistor space, as depicted
in Figure \ref{fig:WilsonLoop}. This mapping essentially swaps edges and vertices. From
a practical standpoint, momentum twistors are very useful because they automatically
encode momentum conservation, but they also have a number of other
important properties. In the context of Yang-Mills
amplitudes, they make the cancellation of spurious poles manifest
and give rise to a geometric interpretation of NMHV amplitudes in terms of the volumes
of polytopes \cite{Hodges:2009hk,ArkaniHamed:2010gg}.

Using momentum twistors, one can define the following invariants of the dual conformal group:
\begin{equation}
\left\langle ijkl\right\rangle =\epsilon_{ABCD}Z_{i}^{A}Z_{j}^{A}Z_{k}^{A}Z_{l}^{A},
\label{4bracket}
\end{equation}
where $\epsilon$ is the Levi-Cevita symbol. We can express the linear dependence of any 5 momentum twistors using this 4-bracket, which can be thought of as a higher-dimensional analogue of the Schouten identity in \eqref{schouten}:
\begin{equation}
Z_{i}\left\langle jklm\right\rangle +{\rm cyclic}=0.\label{plucker}
\end{equation}
The angle-brackets in \eqref{2brackets} can be obtained from momentum twistors using
\eq{
\left\langle ij\right\rangle =I_{AB}Z_{i}^{A}Z_{i}^{B},\,\,\,I_{AB}=\left(\begin{array}{cc}
\epsilon_{\alpha\beta} & 0\\
0 & 0
\end{array}\right),
}
where $I_{AB}$ is called the infinity twistor and breaks dual conformal symmetry. It is also convenient
to define the following 6-brackets \cite{ArkaniHamed:2010gh}:
\begin{align}
\left\langle abc|I|ijk\right\rangle &= \left\langle ab\right\rangle \left\langle cijk\right\rangle +{\rm cyclic}(a,b,c), \nonumber \\
&= \left\langle abci\right\rangle \left\langle jk\right\rangle +{\rm cyclic}(i,j,k),
\label{6bracket1}
\end{align}
\begin{align}
\left\langle a|I|ij|klm\right\rangle &= \left\langle ai\right\rangle \left\langle jklm\right\rangle -\left\langle aj\right\rangle \left\langle iklm\right\rangle, \nonumber \\ 
&= -\left(\left\langle ak\right\rangle \left\langle lmij\right\rangle +{\rm cyclic}(k,l,m)\right),
\label{6bracket2}
\end{align}
where the second lines can be obtained using \eqref{plucker}. These
brackets have an elegant geometrical interpretation in terms of intersections
of lines and planes in momentum twistor space (for more details see \cite{ArkaniHamed:2010gh}).
Moreover, they can be used to express the square brackets in \eqref{2brackets} in terms
of momentum twistors as follows:
\begin{equation}
\left[ij\right]=\frac{\left\langle i-1\,i\,i+1|I|j-1\,j\,j+1\right\rangle }{\left\langle i-1\,i\right\rangle \left\langle i\,i+1\right\rangle \left\langle j-1\,j\right\rangle \left\langle j\,j+1\right\rangle }.\label{square2}
\end{equation}
In general, the numerator in \eqref{square2} has three terms
but there are two cases when it simplifies:
\begin{equation}
\left[i\,i+1\right]=\frac{\left\langle i-1\,i\,i+1\,i+2\right\rangle }{\left\langle i-1\,i\right\rangle \left\langle i\,i+1\right\rangle \left\langle i+1\,i+2\right\rangle },
\label{brsimp1}
\end{equation}
\begin{equation}
\left[i\,i+2\right]=\frac{\left\langle i-1\,i\,i+1\,i+2\right\rangle \left\langle i+3\,i+1\right\rangle +\left\langle i-1\,i\,i+1\,i+3\right\rangle \left\langle i+1\,i+2\right\rangle }{\left\langle i-1\,i\right\rangle \left\langle i\,i+1\right\rangle \left\langle i+1\,i+2\right\rangle \left\langle i+2\,i+3\right\rangle }.
\label{brsimp2}
\end{equation}
Other kinematic invariants such as multi-particle factorisation poles $s_{i_1\dots i_k}$ and spurious poles can also be written in terms of momentum twistor 4-brackets. They likewise take simpler forms when the momentum labels are adjacent.

Finally, we will briefly review the extension to supersymmetric theories.
By analogy to \eqref{eq:region}, we may define fermionic region momenta
\begin{equation}
\left(\theta_{i}-\theta_{i+1}\right)^{\alpha a}=q_{i}^{\alpha a},
\label{theta}
\end{equation}
where the supermomentum $q_{i}$ is defined in \eqref{superm}. Momentum supertwistors
are then defined as $\left(Z_{i}^{A},\chi_{i}^{a}\right)$, where
$\chi_{i}=\theta_{i}\cdot\lambda_{i}$. The supersymmetric extension
of \eqref{line} is then given by
\begin{equation}
\theta_{i}^{\alpha a}=\frac{\lambda_{i}^{\alpha}\chi_{i-1}^{a}-\lambda_{i-1}^{\alpha}\chi_{i}^{a}}{\left\langle i-1i\right\rangle }.\label{superline}
\end{equation}
The following identity for converting fermionic delta functions
to twistor notation will be very useful:
\begin{equation}
\delta^{(0|\mathcal{N})}\left(\left[ii+1\right]\eta_{i+2}+{\rm cyclic}\right)=\frac{\delta^{(0|\mathcal{N})}\left(\left\langle i-1\,i\,i+1\,i+2\right\rangle \chi_{i+3}+{\rm cyclic}\right)}{\left(\left\langle i-1\,i\right\rangle \left\langle i\,i+1\right\rangle \left\langle i+1\,i+2\right\rangle \left\langle i+2\,i+3\right\rangle \right)^{\mathcal{N}}},\label{fermidelta}
\end{equation}
where $\mathcal{N}$ denotes the amount of supersymmetry. We prove
\eqref{fermidelta} in Appendix \ref{app:ferm}. It is also convenient to define the
following object known as an R-invariant:
\begin{equation}
R_{ijk}^{(\mathcal{N})}=\frac{\delta^{(0|\mathcal{N})}\left(\left\langle i\,j-1\,j\,k-1\right\rangle \chi_{k}+{\rm cyclic}\right)}{\left\langle i\,j-1\,j\,k-1\right\rangle \left\langle j-1\,j\,k-1\,k\right\rangle \left\langle j\,k-1\,k\,i\right\rangle \left\langle k-1\,k\,i\,j-1\right\rangle \left\langle k\,i\,j-1\,j\right\rangle },
\label{Rinvariant}
\end{equation}
which first appeared in the context of $\mathcal{N}=4$ SYM \cite{Mason:2009qx}. This
object is invariant under the dual conformal group SU$(4)$ but is
only projectively invariant for $\mathcal{N}=4$, in which case it can
be related to the volume of a polytope in $\mathbb{CP}^{4}$ \cite{ArkaniHamed:2010gg}. This object will
also play a role in supergravity, although its geometrical interpretation
is less clear if $\mathcal{N}\neq4$. 

\section{MHV Examples} \label{mhvsec}

In this section, we will use the on-shell diagram recursion in \eqref{eqn:recursion} to compute MHV amplitudes in $\mathcal{N}=7$ sugra up to 6-points, obtaining Grassmannian integral formulas and expressions in momentum twistor space in agreement with \cite{Hodges:2011wm}. We will see that only $(n-3)!$ diagrams contribute, in contrast to $(n-2)!$ in the $\mathcal{N}=8$ recursion, indicating that the $\mathcal{N}=7$ recursion automatically incorporates the bonus relations.

\subsection{$n=4$}
\label{sec:4ptMHV}
\begin{figure}[H]
\centering
\includegraphics[width=4.25cm]{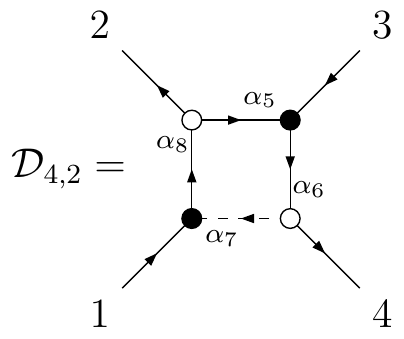}
\caption{On-shell diagram for 4-point amplitude with alternating helicities.}
\label{fig:4ptMHV}
\end{figure}

\begin{figure}[h]
\centering
\includegraphics[width=3cm]{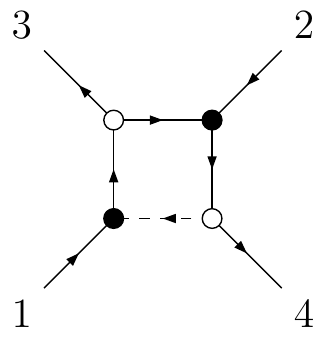}
\caption{On-shell diagram for 4-point amplitude with a split helicity arrangement.}
\label{fig:4ptMHVPerm}
\end{figure}

At four points, only a single on-shell diagram is needed, with no sum over permutations. For $\M_{4,2} (1^-,2^+,3^-,4^+)$, this diagram is shown in Figure \ref{fig:4ptMHV}. If one requires a different helicity arrangement, this can be obtained simply by permuting the external legs. For example, the amplitude $\M_{4,2} (1^-,2^-,3^+,4^+)$ can be obtained from the on-shell diagram in Figure \ref{fig:4ptMHVPerm}. We shall denote individual on-shell diagrams contributing to an amplitude using the symbol $\mathcal{D}$.
Following the algorithm in section \ref{sec:algorithm}, this can be evaluated in terms of edge variables as
\begin{equation}
\mathcal{D}_{4,2} = \frac{1}{\br{41}[14]} \int \prod_{i=5}^8\left(\frac{d\alpha_i}{\alpha_i^2}\right)\frac{1}{\alpha_8\alpha_6} \br{71}[25]\br{53}[47]\,\J_C^3\, \delta^{(4|14)}(C\cdot \ltilde{})\delta^4(\lambda \cdot C^\perp),
\label{4pt}
\end{equation}
where 
\eqs{
C &= \begin{pmatrix}
1	&-\Delta\alpha_8	&0	&-\Delta \alpha_8 \alpha_5 \alpha_6	\\
0	&-\Delta \alpha_6	&1	&-\Delta\alpha_6 \alpha_7\alpha_8	
\end{pmatrix},\\
\J_C &= (1- \alpha_5\alpha_6\alpha_7\alpha_8) = \Delta^{-1},
\label{eqn:4ptCmatrix}
}
and the factor of $\Delta$ comes from a geometric series associated with the closed cycle. The rows of the $C$-matrix are associated with legs 1 and 3. 

When uplifting to a Grassmannian integral there is a Jacobian to transform from an integral over edge variables into one over the entries of the C-matrix:
\begin{equation}
\frac{d^{2\times4}C}{\gl(2)} = \Delta^4\alpha_6^2\alpha_8^2\prod_{i=5}^8d\alpha_i.
\end{equation}
The expression in \eqref{4pt} can then be written as follows:
\eqs{
\mathcal{D}_{4,2} &= \int\frac{d^{2\times 4}C}{\gl(2)}\frac{[23]\br{32}}{\Delta^7\alpha_5\alpha_6^5\alpha_7\alpha_8^5}\delta^{(4|14)}(C\cdot \ltilde{})\delta^4(\lambda \cdot C^\perp),\\
&= \int\frac{d^{2\times 4}C}{\gl(2)}\frac{[23]\br{32}}{(12)(23)(34)(41)(24)^2(31)}\frac{(24)(31)}{(23)(41)}\delta^{(4|14)}(C\cdot \ltilde{})\delta^4(\lambda \cdot C^\perp).
}
This can be simplified by noting that $\br{ij}/(ij)$ and $[ij]/(ij)^\perp$ are independent of $i$ and $j$, where $(ij)^\perp = \varepsilon_{ijkl}(kl)$ is a minor of $C^\perp$. The amplitude then simplifies to
\begin{equation}
\M_{4,2} =\mathcal{D}_{4,2} = \br{13}\int\frac{d^{2\times 4}C}{\gl(2)}\frac{[kl]}{(kl)^\perp}\prod_{i<j}\frac{1}{(ij)}\delta^{(4|14)}(C\cdot \ltilde{})\delta^4(\lambda \cdot C^\perp).\\
\end{equation}

The Grassmannian integral can be evaluated by choosing
\begin{equation}
C = \begin{pmatrix}
\lambda_1	&\lambda_2	&\lambda_3	&\lambda_4\\
\end{pmatrix}.
\end{equation}
This sets $(ij) \to \br{ij}$ and we obtain
\eqs{
\overline{\M}_{4,2} (1^-,2^+,3^-,4^+) &= \br{13} \frac{[23]}{\br{41}}\frac{1}{\prod_{i<j}\br{ij}},
}
where we have factored out the supermomentum delta function to give the quantity $\overline{\M}$, as defined in \eqref{ampbar}. This can be converted into a momentum twistor expression by substituting for the spinor bracket $[23]$ in terms of a twistor 4-bracket according to \eqref{brsimp1}:
\begin{equation}
\overline{\M}_{4,2} (1^-,2^+,3^-,4^+) = \br{13} \frac{\br{1234}}{\prod_i \anbr{i}{i+1} \prod{i<j}\br{ij}}.
\end{equation}

\subsection{$n=5$}
\label{sec:5ptMHV}
\begin{figure}[H]
\centering
\includegraphics[width=5.7cm]{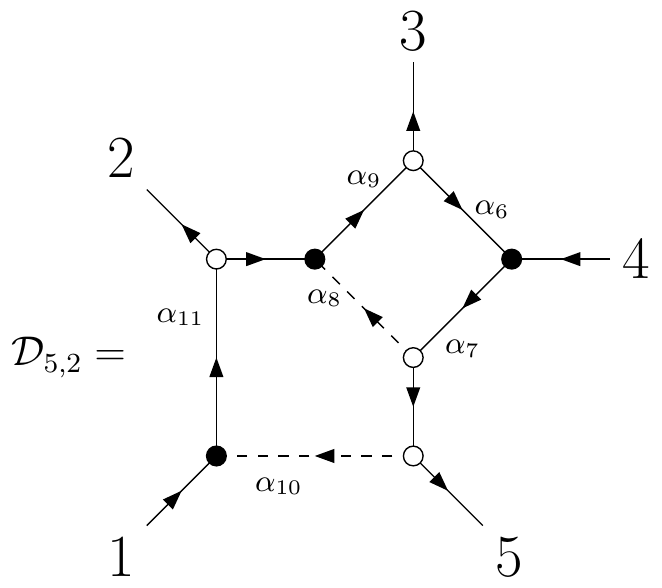}
\caption{On-shell diagram for a 5-point MHV amplitude. To obtain the amplitude, sum over the exchange $2\leftrightarrow3$.}
\label{fig:5ptMHV}
\end{figure}

The 5-point MHV amplitude $\M_{5,2}(1^-,2^+,3^+,4^-,5^+)$ can be obtained from the on-shell diagram in Figure \ref{fig:5ptMHV} after summing over $2\leftrightarrow3$. As we explained in the previous subsection, other helicity arrangements can be obtained by relabelling the diagrams.
Because the diagram has two closed cycles, it is slightly tedious to evaluate the $C$-matrix following the algorithm in section \ref{sec:algorithm}, so we instead use the technique described in section \ref{closedcycle}. First we sum over paths originating from each incoming leg, truncating the path if we reach another incoming leg
\eqs{
\ltilde{1} &= \alpha_{11}\ltilde{2} + \alpha_{11}\alpha_9(\ltilde{3} + \ltilde{6}),\\
\ltilde{4} &= \alpha_7\alpha_8\alpha_9(\ltilde{3} + \ltilde{6}) + \alpha_7\ltilde{5} + \alpha_7\ltilde{10},
\label{eqn:5ptMHVlambdas}
}
where $\ltilde{6}$ and $\ltilde{10}$ are internal spinors adjacent to an incoming leg, which are associated with $\alpha_6$ and $\alpha_{10}$ in Figure \ref{fig:5ptMHV}, respectively. We then we use the fact that all $\tilde{\lambda}$ spinors are proportional at a black vertex to write $\ltilde{6} = \alpha_6\ltilde{4}$ and $\ltilde{10} = \alpha_{10}\ltilde{1}$. Combining these with equation~(\ref{eqn:5ptMHVlambdas}) leads to the $C$-matrix
\eq{
\tilde{C} = \begin{pmatrix}
1	&-\alpha_{11}	&-\alpha_{11}\alpha_9	&-\alpha_{11}\alpha_9\alpha_6	&0\\
-\alpha_7\alpha_{10}	&0	&-\alpha_7\alpha_8\alpha_9	&1-\alpha_7\alpha_8\alpha_9\alpha_6	&-\alpha_7	
\end{pmatrix},}
where the rows are associated with legs 1 and 4. To bring the matrix to a canonical form, we apply a GL$(2)$ transformation given by the inverse of columns 1 and 4. This gives us
\eq{
 G_{\mathrm{fix}} = \Delta_1\Delta_2\begin{pmatrix}
1- \alpha_6\alpha_7\alpha_8\alpha_9	&\alpha_6\alpha_9\alpha_{11}\\
\alpha_{10}\alpha_7	&1\end{pmatrix},}
\eq{
C = G_{\mathrm{fix}}\tilde{C} = \begin{pmatrix}
1	&-\Delta_2 \alpha_{11}	&-\Delta_1\Delta_2\alpha_9\alpha_{11}	&0	&\Delta_1\Delta_2\alpha_9\alpha_{11}\alpha_6\alpha_7\\
0	&-\Delta_1\Delta_2\alpha_7\alpha_{10}\alpha_{11}	&-\Delta_1\alpha_7\alpha_9(\alpha_8 + \alpha_{10}\alpha_{11}\Delta_1\Delta_2)	&1	&-\Delta_1\Delta_2\alpha_7
\end{pmatrix},
}
where $\Delta_1 = (1 - \alpha_6 \alpha_7 \alpha_8 \alpha_9)^{-1}$ and $\Delta_2 = (1 - \Delta_1\alpha_9\alpha_{10}\alpha_{11}\alpha_6\alpha_7)^{-1}$. The Jacobian associated with the closed cycles is then given by
\begin{equation}
\J_C = {\rm det}\left(G_{\rm{fix}}\right)^{-1} = \frac{1}{\Delta_1\Delta_2}.
\end{equation}
The diagram can be written as the following integral over edge variables:
\begin{eqnarray*}
\mathcal{D}_{5,2} &=& \int\prod_{i=6}^{11}\left(\frac{d\alpha_i}{\alpha_i^2}\right)\frac{1}{\alpha_7\alpha_9\alpha_{11}}\frac{\alpha_6\alpha_8\alpha_{10}\br{34}[34][21]}{\alpha_{11}}\J_C^3\,\delta^{(4|14)}(C\cdot \ltilde{})\delta^6(\lambda \cdot C^\perp),\\
&=&\int\frac{d^{2\times5}C}{\gl(2)}\frac{\br{34}[34][21]}{\Delta_1^8\Delta_2^8\alpha_6\alpha_7^6	\alpha_8\alpha_9^5\alpha_{10}\alpha_{11}^7}\delta^{(4|14)}(C\cdot \ltilde{})\delta^6(\lambda \cdot C^\perp),
\label{eqn:5ptMHV_midstep}
\end{eqnarray*}
where the spinor brackets from the decorations have been used to cancel those from the adjacent vertices, and we noted that
\begin{equation}
\frac{d^{2\times5}C}{\gl(2)} = (\Delta_1\Delta_2)^5\alpha_7^3\alpha_9^2\alpha_{11}^3\prod_{i=6}^{11}d\alpha_i.
\end{equation}

To uplift this to a covariant expression, note that
\eq{
\begin{matrix*}[l]
(12) &=& -\Delta_1\Delta_2\alpha_7\alpha_{10}\alpha_{11},\qquad	&	(14) &=& 1,\\
(23) &=& \Delta_1\Delta_2 \alpha_7\alpha_8\alpha_9\alpha_{11},\qquad	&	(25) &=& \Delta_1\Delta_2\alpha_7\alpha_{11},\\
(34) &=& -\Delta_1\Delta_2 \alpha_9\alpha_{11},\qquad	&	(35) &=& \Delta_1\Delta_2 \alpha_7\alpha_9\alpha_{11},\\
(45) &=& \Delta_1\Delta_2 \alpha_6\alpha_7\alpha_9\alpha_{11},\qquad	&	(51) &=& \Delta_1\Delta_2 \alpha_7.\end{matrix*}}
Some products of these are especially useful:
\eqs{
(\Delta_1\Delta_2)^5\alpha_6\alpha_7^4\alpha_8\alpha_9^3\alpha_{10}\alpha_{11}^4 &= \prod_{i=1}^5(i\, i+1),\\
\Delta_1\Delta_2 &= \frac{(15)(34)}{(35)(14)},\\
\alpha_{11} &= \frac{(25)}{(51)},\\
(\Delta_1\Delta_2 \alpha_7\alpha_9\alpha_{11})^2 &= \frac{(35)^2}{(14)}.
}
Recalling that $\br{ij}/(ij)$ is independent of $i$ and $j$, we finally get
\begin{equation}
\mathcal{D}_{5,2} = \br{14}\int\frac{d^{2\times5}C}{\gl(2)}[12][34](24)(13)\prod_{i<j}\frac{1}{(ij)}\delta^{(4|14)}(C\cdot \ltilde{})\delta^6(\lambda \cdot C^\perp).
\end{equation}
The 5-point amplitude can then be recovered by summing the above formula over the permutation $2 \leftrightarrow 3$:
\begin{equation}
\mathcal{M}_{5,2}(1^-,2^+,3^+,4^-,5^+) = \br{14}\int\frac{d^{2\times5}C}{\gl(2)}([12][34](24)(13) - [13][24](34)(12))\frac{\delta^{(4|14)}(C\cdot \ltilde{})\delta^6(\lambda \cdot C^\perp)}{\prod_{i<j}(ij)}.
\label{eqn:5ptMHVAmp}
\end{equation}

The Grassmannian integral formula in \eqref{eqn:5ptMHVAmp} can be evaluated by setting the columns of $C$ to $\lambda_i$ giving
\begin{equation}
\overline{\M}_{5,2}(1^-,2^+,3^+,4^-,5^+) = \br{14} \frac{[12][34]\br{24}\br{13} - [13][24]\br{12}\br{34}}{\prod_{i<j}\br{ij}}.
\end{equation}
Note that $[12][34]\br{24}\br{13} - [13][24]\br{12}\br{34} = 4i \epsilon_{\mu\nu\rho\sigma}p_1^\mu p_2^\nu p_3^\rho p_4^\sigma$ is permutation-invariant on support of momentum conservation.  Hence, we can equivalently write it as $[23][45]\br{35}\br{24} - [24][35]\br{23}\br{45}$. To convert the numerator to momentum twistor notation, let us consider the following quantity, first defined in \cite{Hodges:2011wm}:
\eqs{
N_5 &=  \big([23][45]\br{35}\br{24} - [24][35]\br{23}\br{45}\big)\br{12}\br{23}\br{34}\br{45}\br{51}, \\
&=\br{1234}\br{2345}\br{51} - \br{1234}\br{3451}\br{25} - \br{5123}\br{2345}\br{14}.
\label{N5}
}
This quantity will also play a role for non-MHV amplitudes. To prove the second equality, use \eqref{brsimp1}, \eqref{brsimp2}, and \eqref{plucker}:
\eqs{
N_5 &= \frac{\br{1234}\br{3451}\br{35}\br{24}}{\br{34}} - \frac{(\br{1234}\br{53} + \br{5123}\br{43})(\br{4512}\br{43} + \br{3451}\br{42})}{\br{34}},\\
&= -(\br{1234}\br{4512}\br{35} + \br{5123}\br{4512}\br{34} + \br{5123}\br{3451}\br{24}).
}
Equation \eqref{N5} then follows from noting that $N_5$ is invariant under cyclic permutations.
Hence, the 5-point MHV amplitude has the following form in momentum twistor space:
\begin{equation}
\overline{\M}_{5,2}(1^-,2^+,3^+,4^-,5^+) = \br{14} \frac{\br{1234}\br{2345}\br{51} - \br{1234}\br{3451}\br{25} - \br{2345}\br{5123}\br{14}}{\prod_i\anbr{i}{i+1} \prod_{i<j}\br{ij}}.
\end{equation}

\subsection{$n = 6$}
\begin{figure}[H]
\centering
\includegraphics[width=6.5cm]{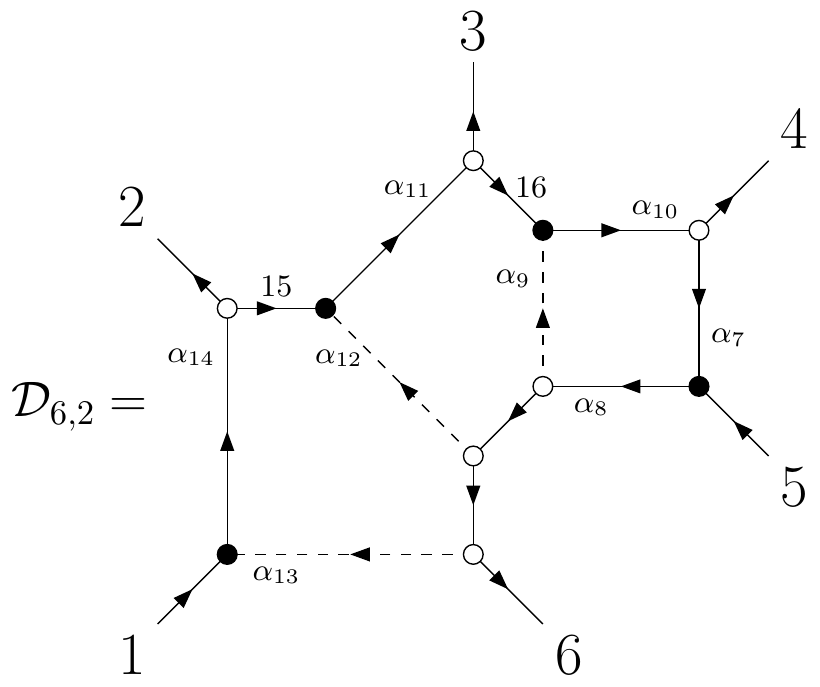}
\caption{On-shell diagram for a 6-point MHV amplitude. This diagram needs to be summed over all permutations of $\left\{ 2,3,4\right\} $.}
\label{fig:6ptMHV}
\end{figure}

The 6-point MHV amplitude $\M_{6,2}(1^-,2^+,3^+,4^+,5^-,6^+) $ can be obtained by summing the diagram shown in Figure \ref{fig:6ptMHV} over permutations of legs $\left\{ 2,3,4\right\} $. As before, other helicity arrangements can be obtained by relabelling.
Using the technique described in section \ref{closedcycle}, we obtain the C-matrix
\begin{equation}
\tilde{C} = \begin{pmatrix}
1	&-\alpha_{14}	&-\alpha_{11}\alpha_{14}	&-\alpha_{10}\alpha_{11}\alpha_{14}	&-\alpha_7\alpha_{10}\alpha_{11}\alpha_{14}	&0\\
-\alpha_8\alpha_{13}	&0	&-\alpha_8\alpha_{11}\alpha_{12}	&-\alpha_8\alpha_{10}(\alpha_9 + \alpha_{11}\alpha_{12})	&1 -\alpha_7\alpha_8\alpha_{10}(\alpha_9 + \alpha_{11}\alpha_{12})	&-\alpha_8
\end{pmatrix}.
\end{equation}
We then gauge-fix using
\eqs{
G_{\mathrm{fix}} &= \Delta\begin{pmatrix}
1-\alpha_7\alpha_8\alpha_{10}(\alpha_9+\alpha_{11}\alpha_{12})	&\alpha_7\alpha_{10}\alpha_{11}\alpha_{14}\\
\alpha_8\alpha_{13}	&1
\end{pmatrix},
}
and the Jacobian associated with closed cycles is
\eqs{
\mathcal{J}_C &= \Delta^{-1} = \left(1 - \alpha_{10}\alpha_7\alpha_8\left(\alpha_9 +\alpha_{11}(\alpha_{12} + \alpha_{13}\alpha_{14})\right)\right).
}
This leads to the $C$-matrix
\eqs{
C = G_{\mathrm{fix}}\tilde{C} &= \bigg(\begin{matrix}
1	&-\Delta\alpha_{14}\left(1-\alpha_7\alpha_8\alpha_{10}(\alpha_9+\alpha_{11}\alpha_{12})\right)	&-\Delta\alpha_{11}\alpha_{14}(1-\alpha_7\alpha_8\alpha_9\alpha_{10})\\
0	&-\Delta\alpha_8\alpha_{13}\alpha_{14}&-\Delta\alpha_8\alpha_{11}(\alpha_{12}+\alpha_{13}\alpha_{14})	
\end{matrix}\\
&\qquad\qquad\qquad
\begin{matrix}
-\Delta\alpha_{10}\alpha_{11}\alpha_{14}	&0	&-\Delta\alpha_7\alpha_8\alpha_{10}\alpha_{11}\alpha_{14}\\
-\Delta\alpha_8\alpha_{10}\left(\alpha_9+\alpha_{11}(\alpha_{12}+\alpha_{13}\alpha_{14})\right)	&1	&-\Delta\alpha_8
\end{matrix}\bigg).
}
The bracket factors from the vertices are
\eqs{
[47] &= \alpha_7[45],\\
\br{75} &= \br{45},\\
\sqbr{2}{15} &= \frac{1}{\alpha_{14}}[21],\\	
\sqbr{3}{16} &= \frac{1}{\alpha_{11}\alpha_{14}}\left([31] - \alpha_{14}[32]\right),\\& =-\frac{1}{(36)}\left([31](16)+ [32](26)\right).
}
Moreover the following expressions relating edge variables to minors are useful for obtaining a Grassmannian integral formula:
\eqs{
\frac{d^{2\times6}C}{\gl(2)} &= \Delta^6\alpha_8^4\alpha_{10}^2\alpha_{11}^3\alpha_{14}^4 \prod_{i=7}^{14} d\alpha_i,\\
\prod_{i=1}^6(i\,i+1) &= \Delta^6\alpha_7\alpha_8^5\alpha_9\alpha_{10}^3\alpha_{11}^4\alpha_{12}\alpha_{13}\alpha_{14}^5,\\
(16) &= -\Delta \alpha_8,\\
(26) &= \Delta \alpha_8\alpha_{14},\\
(36) &=\Delta \alpha_8\alpha_{11}\alpha_{14},\\
(45) &= -\Delta \alpha_{10}\alpha_{11}\alpha_{14},\\
(46) &=\Delta \alpha_8\alpha_{10}\alpha_{11}\alpha_{14}.}

After collecting all the terms, the on-shell diagram in Figure \ref{fig:6ptMHV} evaluates to
\begin{eqnarray*}
\mathcal{D}_{6,2} &=& \int\prod^{14}_{i=7}\left(\frac{d\alpha_i}{\alpha_i^2}\right)\frac{\alpha_9\alpha_{12}\alpha_{13}\alpha_7}{\alpha_8\alpha_{10}\alpha_{11}\alpha_{14}^2}\frac{[45]\br{45}[12]\left([31](16) + [32](26)\right)}{\ (36)}\J_C^3\delta^{(4|14)}(C\cdot \ltilde{})\delta^8(\lambda \cdot C^\perp),\\
&=&\int\frac{d^{2\times6}C}{\gl(2)}\frac{\delta^{(4|14)}(C\cdot \ltilde{})\delta^8(\lambda \cdot C^\perp)}{\Delta^9\alpha_7\alpha_8^7\alpha_9\alpha_{10}^5\alpha_{11}^6\alpha_{12}\alpha_{13}\alpha_{14}^8}\frac{[45]\br{45}[12]\left([31](16) + [32](26)\right)}{(36)},\\
&=& \int \frac{d^{2\times6}C}{\gl(2)}\frac{\delta^{(4|14)}(C\cdot \ltilde{})\delta^8(\lambda \cdot C^\perp)}{\prod_{i=1}^6(i\,i+1)}\frac{[45]\br{45}[12]\left([31](16) + [32](26)\right)}{(36)(45)(64)(26)}.\\
\end{eqnarray*}
The Grassmannian integral can then be evaluated to give the spinorial expression
\begin{equation}
\overline{\mathcal{D}}_{6,2}=\br{15} \prod_{i<j}\,\frac{1}{\br{ij}}[45][12][3|1+2|6\rangle \br{25}\br{35}\br{24}\br{13}\br{14}.
\label{eqn:6ptMHVTerm}
\end{equation}
Finally, the six-point amplitude can be recovered by summing over the six permutations of legs 2, 3 and 4:
\begin{equation}
\overline{\M}_{6,2}(1^-,2^+,3^+,4^+,5^-,6^+) = \sum_{\mathcal{P}(2,3,4)}\overline{\mathcal{D}}_{6,2},
\label{eqn:6ptMHVresult}
\end{equation}
which is the BGK form of the MHV gravity amplitude \cite{Berends:1988zp}.

Equation (\ref{eqn:6ptMHVresult}) can be simplified by using momentum conservation to eliminate the square brackets $\left\{ \left[23\right],\left[34\right],\left[42\right],\left[56\right],\left[61\right],\left[15\right]\right\}$, and then using the Schouten identity to eliminate the corresponding angle brackets. These are the brackets that transform into each other (or not at all) when we carry out the permutation sum. We are left with the following sum:
\eq{
\overline{\M}_{6,2}(1^-,2^+,3^+,4^+,5^-,6^+) = \br{15}\sum_{\mathcal{P}(2,3,4)}\frac{[12][53][64]\br{13}\br{14}\br{52}\br{54}\br{62}\br{63}}{\prod_{i<j}\br{ij}}. 
}
Each of the six terms can be translated into a product of 6-brackets using \eqref{square2}. However, if we first exchange $3\leftrightarrow6$ (using the permutation symmetry of the amplitude), this removes all square brackets of the form $[ii{+}2]$ and also ensures maximum cancellation of angle 2-brackets between numerator and denominator. This leads to the same compact twistor expression found in \cite{Hodges:2011wm}, which can be expressed as
\begin{equation}
\overline{\M}_{6,2}(1^-,2^+,3^+,4^+,5^-,6^+) = \br{15}\frac{N_6}{\prod_i\anbr{i}{i{+}1}\prod_{i<j}\br{ij}},
\label{6ptmhvtwist}
\end{equation}
with
\begin{eqnarray*}
N_6&=& \br{123|I|456}\br{234|I|561}\br{345|I|612}\\
&&+\,\br{123|I|456}\br{5612}\br{2345}\br{14}\br{36}\\
&&+\,\br{234|I|561}\br{6123}\br{3456}\br{25}\br{41}\\
&&+\,\br{345|I|612}\br{1234}\br{4561}\br{36}\br{52}\\
&&+\, \br{1234}\br{3456}\br{5612}\br{14}\br{25}\br{36}\\
&&+\,\br{2345}\br{4561}\br{6123}\br{25}\br{36}\br{41}.
\end{eqnarray*}
In \cite{Hodges:2011wm}, Hodges conjectured that the structure of \eqref{6ptmhvtwist} can be extended to all gravitational MHV amplitudes.

\section{NMHV Examples} \label{nmhvsec}
In this section, we will use the on-shell diagram recursion to compute NMHV amplitudes in $\mathcal{N}=7$ sugra. As a warm-up, we first consider the 5-point NMHV amplitude. Although this is just the parity conjugate of an MHV amplitude, converting it to momentum twistor space reveals interesting structure which extends to higher points, notably R-invariants analogous to the building blocks for non-MHV amplitudes in $\mathcal{N}=4$ SYM. We then go on to compute the 6-point NMHV amplitude. Unlike MHV amplitudes, the on-shell diagrams for non-MHV amplitudes correspond to residues of top-forms in the Grassmannian (this was previously observed in $\mathcal{N}=4$ SYM in \cite{ArkaniHamed:2009dn,ArkaniHamed:2009dg,Nandan:2009cc} and $\mathcal{N}=8$ sugra in \cite{Farrow:2017eol}). 

Another difference compared to MHV amplitudes is that we do not use globally defined momentum twistors to describe non-MHV amplitudes above 5-points. In particular, for the 6-point NMHV amplitude we split the on-shell diagrams into three sets and define momentum twistors with respect to a different ordering of external momenta in each set, analogous to defining local coordinates on different patches of a manifold. Although this  approach is unconventional, it has two important pay-offs. Firstly, the amplitudes can be expressed in terms of R-invariants similar to those found a 5-points. Secondly, the cancellation of spurious poles becomes completely transparent.

\subsection{$n=5$}
\label{sec:5ptNMHV}
\begin{figure}[H]
\centering
\includegraphics[width=6cm]{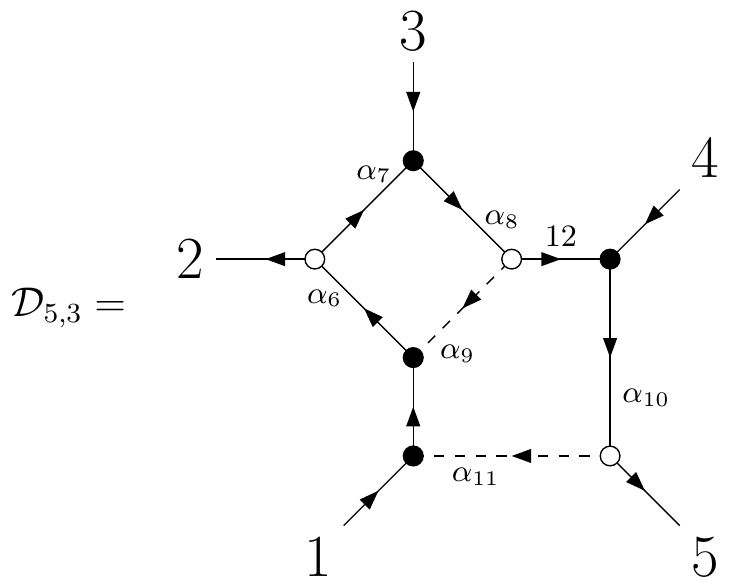}
\caption{On-shell diagram for a 5-point NMHV amplitude. The full amplitude is recovered by summing over $3\leftrightarrow4$.}
\label{fig:5ptNMHV}
\end{figure}

We will begin by computing the 5-point NMHV amplitude $\M_{5,3}(1^-,2^+,3^-,4^-,5^+)$, which can be obtained from the on-shell diagram in Figure \ref{fig:5ptNMHV} by summing over $3\leftrightarrow4$.
Using the technique in section \ref{closedcycle} we obtain the C-matrix
\begin{equation}
\tilde{C} = \begin{pmatrix}
1	&-\alpha_6	&-\alpha_6\alpha_7	&0	&0\\
0	&-\alpha_6\alpha_8\alpha_9	&1 - \alpha_6\alpha_7\alpha_8\alpha_9	&-\alpha_8	&0\\
-\alpha_{10}\alpha_{11}	&0	&0	&1	&-\alpha_{10}
\end{pmatrix}.
\end{equation}
This can be put in canonical form using the GL$(3)$ transformation
\begin{align}
\ G_{\mathrm{fix}} &= \Delta
\begin{pmatrix}
1 - \alpha_6\alpha_7\alpha_8\alpha_9	&\alpha_6\alpha_7	&\alpha_6\alpha_7\alpha_8\\
\alpha_8\alpha_{10}\alpha_{11}	&1	&\alpha_8\\
\alpha_{10}\alpha_{11}(1 - \alpha_6\alpha_7\alpha_8\alpha_9)	&\alpha_6\alpha_7\alpha_{10}\alpha_{11}	&1 - \alpha_6\alpha_7\alpha_8\alpha_9
\end{pmatrix},
\end{align}
where the Jacobian associated with closed cycles is
\begin{equation}
\J_C=\Delta^{-1} = \big(1 - \alpha_6\alpha_7\alpha_8(\alpha_9 + \alpha_{10}\alpha_{11})\big).
\end{equation}
The gauge-fixed $C$-matrix is then given by
\eq{
C = \begin{pmatrix}
1	&-\Delta\alpha_6	&0	&0	&-\Delta\alpha_6\alpha_7\alpha_8\alpha_{10}\\
0	&-\Delta\alpha_6\alpha_8(\alpha_9 + \alpha_{10}\alpha_{11})	&1	&0	&-\Delta\alpha_8\alpha_{10}\\
0	&-\Delta\alpha_6\alpha_{10}\alpha_{11}	&0	&1	&-\Delta\alpha_{10}(1 - \alpha_6\alpha_7\alpha_8\alpha_9)
\end{pmatrix}.
}
The bracket factors from the vertices are
\eqs{
\anbr{13}{4} &= \frac{1}{\alpha_{10}}\br{54}\\
[27] &= \alpha_7[23]\\
\br{73} &= \br{23}.
}

To derive a Grassmannian integral formula the following relations are useful:
\eqs{
\frac{d^{3\times5}C}{\gl(3)} &= \Delta^5\alpha_6^3\alpha_8^2\alpha_{10}^3\prod_{i=7}^{11}\alpha_i, \\
\prod_{i=1}^5(i\,i+1\,i+2) &= \Delta^5\alpha_6\alpha_7\alpha_8^2\alpha_{10}^3,\\
\frac{(145)(234)}{(245)} &= -\Delta,\\
\frac{(235)}{(234)} &= -\alpha_{10},\\
(245) &= -\Delta\alpha_6\alpha_8\alpha_{10}.
}
Using these relations, the diagram in Figure \ref{fig:5ptNMHV} evaluates to 
\eqs{
\mathcal{D}_{5,3} &= \int\left(\prod_{i=6}^{11}\frac{d\alpha_i}{\alpha_i^2}\right)\frac{\delta^{(6|21)}(C\dotprod\ltilde{})\delta^4(\lambda\dotprod C^\perp)}{\alpha_6\alpha_8\alpha_{10}}\frac{\alpha_7\alpha_9\alpha_{11}}{\alpha_{10}}\br{54}[23]\br{23}\J_C^3,  \\ 
&= \int\frac{d^{3\times5}C}{\gl(3)}\frac{\br{54}[23]\br{23}}{\Delta^8\alpha_6^6\alpha_7^2\alpha_8^5\alpha_9\alpha_{10}^7\alpha_{11}}\delta^{(6|21)}(C\dotprod\ltilde{})\delta^4(\lambda\dotprod C^\perp), \\
&= \int\frac{d^{3\times5}C}{\gl(3)}\delta^{(6|21)}(C\dotprod\ltilde{})\delta^4(\lambda\dotprod C^\perp)\frac{\br{54}[23]\br{23}}{\prod_{i=1}^5(i\,i+1\,i+2)(235)(145)(245)},  \\
&=[25]\int\frac{d^{3\times5}C}{\gl(3)}\delta^{(6|21)}(C\dotprod\ltilde{}|\eta)\delta^4(\lambda\dotprod C^\perp)\frac{\br{54}\br{23}(124)(135)}{\prod_{i<j<k}(ijk)},
\label{5ptnmhvgrass}
}
where we have used that $[23]/[25] = (145)/(134)$. This relation can be derived by noting that for the canonical choice
\begin{equation}
C=\left(\begin{array}{ccccc}
1 & c_{12} & 0 & 0 & c_{15}\\
0 & c_{32} & 1 & 0 & c_{35}\\
0 & c_{42} & 0 & 1 & c_{45}
\end{array}\right),
\end{equation}
the bosonic delta functions $\delta^6\left(C\cdot\tilde{\lambda}\right)$ imply that $c_{i2}=\left[5i\right]/\left[25\right]$ and $c_{i5}=\left[i2\right]/\left[25\right]$, so $(ijk)=\epsilon_{ijkmn}[lm]/[25]$. Plugging this into \eqref{5ptnmhvgrass} and summing over $3\leftrightarrow 4$ finally gives
\begin{equation}
\overline{\M}_{5,3} = \frac{[25]}{\prod_{i<j}[ij]}\left([35][24]\br{23}\br{54} - [45][23]\br{24}\br{53}\right)\delta^{(0|7)}\left(\frac{[51]\eta_{2}+{\rm cyclic}}{\left\langle 34\right\rangle }\right).
\label{eqn:5ptNMHV}
\end{equation}

If we pull out a helicity-dependent prefactor, \eqref{eqn:5ptNMHV} can alternatively be written as
\begin{equation}
\overline{\M}_{5,3}(1^-,2^+,3^-,4^-,5^+) = [25]\frac{1}{\br{12}\br{23}\br{34}\br{45}\br{51}}\,\hat{M}_5,
\label{5ptnmhvamp}
\end{equation}
where $\hat{M}_5$ is little-group invariant which has the following form in momentum twistor space:
\begin{equation}
\hat{M}_5 =  \frac{N_5 R^{(7)}_{135}}{D_5},
\label{Mhat}
\end{equation}
where $R^{(7)}$ is an $\mathcal{N}=7$ R-invariant defined in \eqref{Rinvariant}, $N_5$ is defined in \eqref{N5}, and 
\begin{equation}
D_5=\prod_{i=1}^5\langle i-2\,i-1\,i |I|i\,i+1\,i+2\rangle.
\end{equation}
Note that $D_5$ is proportional to the product of five spinor brackets $[ii{+}2]$ using \eqref{square2} and \eqref{brsimp2}. Neither $R^{(7)}$ nor $D_5$ are permutation invariant but their ratio should nevertheless have the $S_5$ symmetry since we pulled out the helicity-dependent part of the amplitude in \eqref{5ptnmhvamp}. In $\mathcal{N}=4$ SYM, R-invariants form the building blocks for all tree-level non-MHV amplitudes, so it is interesting to see them appear in sugra amplitudes. In \cite{Drummond:2009ge}, $\mathcal{N}=8$ sugra amplitudes were constructed by squaring $\mathcal{N}=4$ R-invariants, but in this paper we define R-invariants which are more intrinsic to sugra. Note that both $R^{(4)}$ and $R^{(7)}$ are invariant under the dual conformal group SL$(4)$, but $R^{(4)}$ have additional GL$(1)$ symmetry which makes them projective and leads to an elegant geometric interpretation of NMHV amplitudes in $\mathcal{N}=4$ SYM as volumes of polytopes in $\mathbb{CP}^{4}$ \cite{ArkaniHamed:2010gg}. Note that $\hat{M}_5$ in \eqref{Mhat} is also GL$(4)$ invariant, so it would be interesting explore its geometric interpretation. 

Writing the remaining spinor bracket in \eqref{5ptnmhvamp} in terms of a twistor bracket using \eqref{square2}, we get
\begin{equation}
\overline{\M}_{5,3}(1^{-},2^{+},3^{-},4^{-},5^{+})=\frac{R_{135}^{(7)}}{\left[\mathrm{PT}(5)\right]^{2}}\frac{\br{123|I|451}\br{34}N_{5}}{D_{5}},
\label{eqn:5ptNMHVwHelicity}
\end{equation}
where 
\begin{equation}
\left[\mathrm{PT}(n)\right]=\Pi_{i=1}^{n}\left\langle ii+1\right\rangle 
\end{equation}
and $n+1$ is identified with particle $1$. We will find similar structure for 6-point NMHV amplitudes.

\subsection{$n=6$} \label{6ptnmhvos}

In this subsection, we will consider the amplitude with alternating helicities $\M_{6,3}(1^-,2^+,3^-,4^+,5^-,6^+)$. Any other can then be obtained by a relabelling. Using the recursion in \eqref{eqn:recursion}, this amplitude can be obtained from four on-shell diagrams summed over permutations to give a total of 13 terms. It is then natural to combine them into 9 terms, each with a common pole of the form  $s_{ijk} = (p_i + p_j + p_k)^2$. In this way we obtain a superamplitude whose graviton component is equivalent to the spinorial expression obtained in \cite{Hodges:2011wm} up to a relabelling. In the next subsection we obtain a new formula for the 6-point NMHV amplitude in terms of momentum twistors defined with respect to different orderings of the external momenta, which reveals surprising mathematical structure and provides a systematic understanding of spurious pole cancellation.  

Since $\mathcal{N}=7$ on-shell diagrams are labelled with arrows encoding the helicities of the superfields, we will sometimes have to use non-cyclic labels for the external legs even before we sum over permutations. Our description of how to evaluate the on-shell diagrams will be more schematic in this section. For more details, see Appendix \ref{6ptdetail}.

\subsubsection*{3+5 and 5+3 Diagrams}
\begin{figure}[H]
\centering
\includegraphics[width=5.2cm]{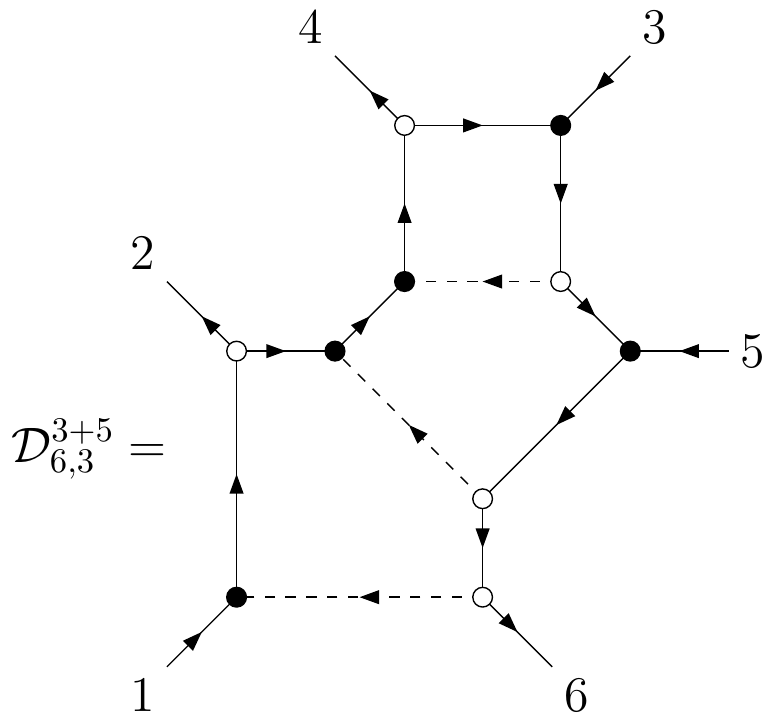}
\caption{On-shell diagram combining a 3-point $\overline{\rm{MHV}}$ amplitude with a 5-point $\overline{\rm{MHV}}$ amplitude. This diagram needs to be summed over the permutations $2\leftrightarrow4$ and $3\leftrightarrow5$.}
\label{fig:6ptNMHV_3+5}
\end{figure}

\begin{figure}[H]
\centering
\includegraphics[width=5.2cm]{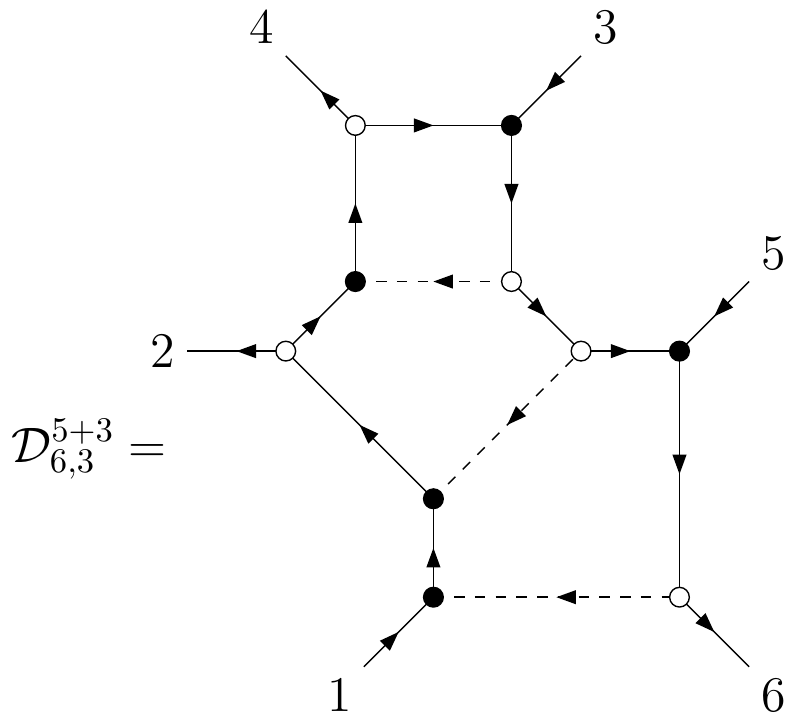}
\caption{On-shell diagram combining a 5-point MHV amplitude with a 3-point MHV amplitude. This diagram needs to be summed over the permutations $2\leftrightarrow4$ and $3\leftrightarrow5$.}
\label{fig:6ptNMHV_5+3}
\end{figure}

We begin with the diagrams in Figures \ref{fig:6ptNMHV_3+5} and \ref{fig:6ptNMHV_5+3}, which encode factorisations into 3-point and 5-point subamplitudes. 
Using the methods described in section \ref{osdsec}, they can be written compactly as
\eqs{
\mathcal{D}^{3+5}_{6,3} &=\res_{(612) = 0}\int d^{3\times 6}\Omega_7\frac{\br{34}[34]\br{56}[12]}{(346)(256)(356)}\frac{(123)}{(124)},\\
\mathcal{D}^{5+3}_{6,3} &= \res_{(234) = 0} \int d^{3\times 6}\Omega_7\frac{\br{34}[34]\br{56}[12]}{(346)(256)(356)}\frac{(123)}{(124)},
\label{35grass}
}
where 
\begin{equation}
d^{3\times 6}\Omega_7 = \frac{d^{3\times6}C}{\gl(3)}\frac{\delta^{(6|21)}(C\cdot \ltilde{}|\eta)\delta^{(6)}(\lambda\cdot C^\perp)}{\prod_{i=1}^6(i\,i+1\,i+2)}.
\end{equation}
These diagrams need to be summed over the permutations $2\leftrightarrow4$ and $3\leftrightarrow 5$. In each case it can be seen that one of these permutations only affects the integrand whilst the other also changes which residue we take.

The Grassmannian integrals in \eqref{35grass} can be evaluated using a clever choice of gauge \cite{Arkani-Hamed:2016byb}. For example, to evaluate $\mathcal{D}^{3+5}_{6,3} $ we multiply by $(612)$ and choose 
\begin{equation}
C = \begin{pmatrix}
\lambda_1	&\lambda_2	&\lambda_3	&\lambda_4	&\lambda_5	&\lambda_6\\
0	&0	&[45]	&[53]	&[34]	&0
\end{pmatrix}.
\end{equation}
Calculating the required minors and substituting into the first line of \eqref{35grass} then gives
\begin{equation}
\overline{\mathcal{D}}^{3+5}_{6,3} = \frac{\br{34}\br{56}[12]\,\delta^{(0|7)}\left([45]\eta_3 + [53]\eta_4 + [34]\eta_5\right)}{s_{345}\br{12}[35][34]\br{61}\br{26}[3|4+5|6\rangle[4|5+3|6\rangle[5|3+4|6\rangle[5|3+4|2\rangle},
\end{equation}
where $[i|j+k|l\rangle = [ij]\br{jl} +[ik]\br{kl}$. Terms of this form correspond to spurious poles which must cancel out in the amplitude. The full expression containing an $s_{345}$ pole is obtained by summing over the exchange $3\leftrightarrow5$:
\begin{equation}
\overline{\mathcal{D}}_{6,3}^{(345)} = \frac{[12]\,\delta^{(0|7)}\left([45]\eta_3 + [53]\eta_4 + [34]\eta_5\right)}{s_{345}\br{12}\br{26}\br{61}[35][3|4+5|6\rangle[4|5+3|6\rangle[5|3+4|6\rangle}\left(\frac{\br{34}\br{56}}{[34][5|3+4|2\rangle} - \frac{\br{54}\br{36}}{[54][3|5+4|2\rangle}\right),
\label{eqn:345Term}
\end{equation}
where the sign which comes from exchanging of $3\leftrightarrow5$ in the fermionic delta function is cancelled by the sign which comes from anticommuting the corresponding $\Phi^-$ superfields. The term with an $s_{235}$ pole can be obtained by applying the permutation $2\leftrightarrow4$ to \eqref{eqn:345Term}. 

Similarly, for $\mathcal{D}^{5+3}_{6,3} $, we choose
\begin{equation}
C = \begin{pmatrix}
\lambda_1	&\lambda_2	&\lambda_3	&\lambda_4	&\lambda_5	&\lambda_6\\
[56]	&0	&0	&0	&[61]	&[15]\end{pmatrix},
\end{equation}
which gives
\begin{equation}
\overline{\mathcal{D}}^{(5+3)}_{6,3} = \frac{[34]\br{56}[12]\,\delta^{(0|7)}\left([56]\eta_1 + [61]\eta_5 + [15]\eta_6\right)}{s_{561}[61][15][56] \br{43}\br{24}[1|5+6|2\rangle[1|5+6|3\rangle[1|5+6|4\rangle[5|6+1|2\rangle}.
\end{equation}
The full expression with an $s_{561}$ pole is then obtained by summing over the permutation $2\leftrightarrow4$ and is given by
\begin{equation}
\overline{\mathcal{D}}^{(561)}_{6,3} = \frac{\br{56}\,\delta^{(0|7)}\left([56]\eta_1 + [61]\eta_5 + [15]\eta_6\right)}{s_{561}[61][15][56]\br{24}[1|5+6|2\rangle[1|5+6|3\rangle[1|5+6|4\rangle}\left(\frac{[12][34]}{ \br{43}[5|6+1|2\rangle} - \frac{[14][32]}{\br{23}[5|6+1|4\rangle}\right).
\label{eqn:561Term}
\end{equation}
The term with a pole in $s_{361}$ is then obtained by applying the permutation $3\leftrightarrow5$.

\subsubsection*{4+4 Diagrams}
\begin{figure}[H]
\centering
\includegraphics[width=7cm]{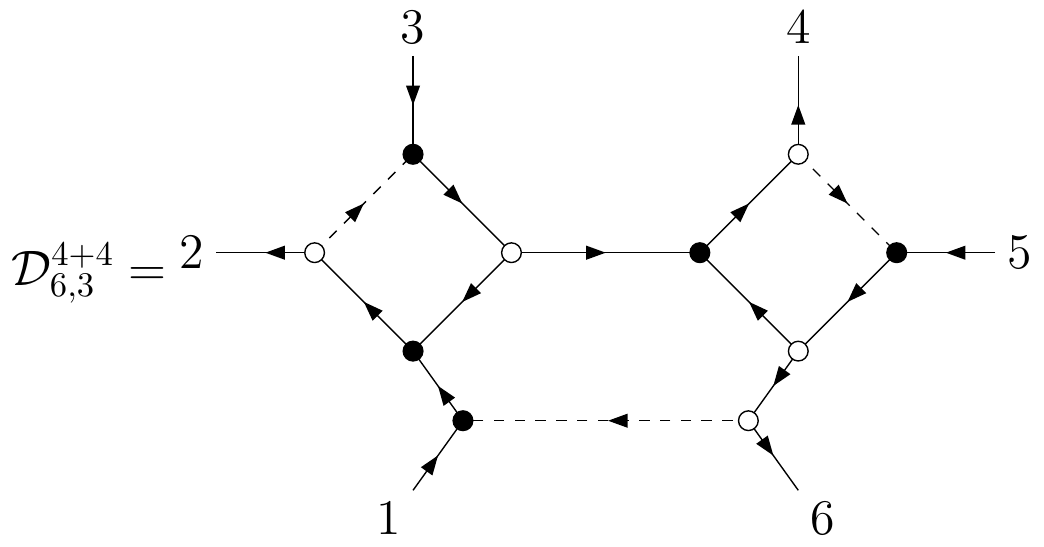}
\caption{On-shell diagram combining two 4-point amplitudes with alternating helicities. This diagram needs to be summed over the permutations $2\leftrightarrow4$ and $3\leftrightarrow5$.}
\label{fig:6ptNMHV_4+4}
\end{figure}

\begin{figure}[h]
\centering
\includegraphics[width=7cm]{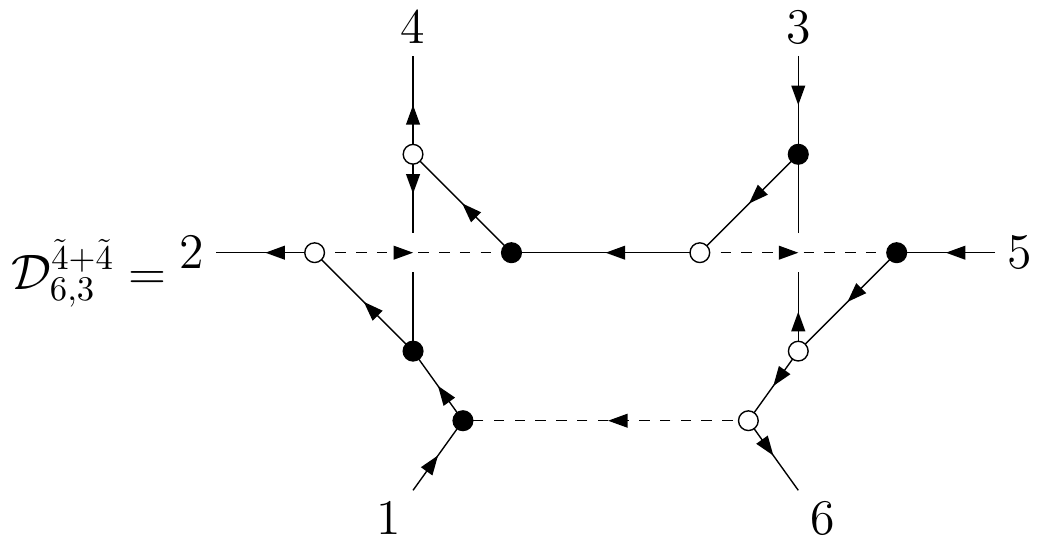}
\caption{On-shell diagram combining two 4-point amplitudes with split helicities. This diagram does not require any permutation sums since it is invariant under $2\leftrightarrow4$ and $3\leftrightarrow5$.}
\label{fig:6ptNMHV_4+4Twisted}
\end{figure}

These diagrams encode factorisations into 4-point subamplitudes. In the $\N=7$ formalism, there are two inequivalent types in which the 4-point amplitudes either have negative helicities opposite or adjacent to each other, as depicted in Figures \ref{fig:6ptNMHV_4+4} and \ref{fig:6ptNMHV_4+4Twisted}, respectively.
Using the methods described in section \ref{osdsec}, the diagram in Figure \ref{fig:6ptNMHV_4+4} evaluates to
\begin{equation}
\mathcal{D}^{(4+4)}_{6,3} = \res_{(456) = 0}\int d^{3\times 6}\Omega_7\frac{[13][45]\br{23}\br{46}}{(236)(246)^2}.
\end{equation}
To find a spinor expression for this diagram, we choose
\begin{equation}
C = \begin{pmatrix}
\lambda_1	&\lambda_2	&\lambda_3	&\lambda_4	&\lambda_5	&\lambda_6\\
[23]	&[31]	&[12]	&0	&0	&0
\end{pmatrix},
\end{equation}
which gives
\begin{eqnarray}
\overline{\mathcal{D}}^{(4+4)}_{6,3} &\equiv& \overline{\mathcal{D}}^{(123)}_{6,3}, \nonumber \\
&=& -\frac{[13][45]\br{23}\br{46}\,\delta^{(0|7)}\left([23]\eta_1 + [31]\eta_2 + [12]\eta_3\right)}{s_{123}[1|2+3|4\rangle[12][45]\br{56}[23][3|1+2|6\rangle[31]^2\br{46}^2[1|2+3|6\rangle}, \nonumber \\
&=& \frac{\br{23}[45]\,\delta^{(0|7)}\left([23]\eta_1 + [31]\eta_2 + [12]\eta_3\right)}{s_{123}[12][23][31]\br{46}\br{45}\br{56}[1|2+3|4\rangle[1|2+3|6\rangle[3|1+2|6\rangle}. \label{eqn:123Term}
\end{eqnarray}

The final diagram in Figure \ref{fig:6ptNMHV_4+4Twisted} is non-planar. This is a consequence of using the BCFW bridge in \eqref{eqn:decorationdefn} with its fixed helicity assignments . The diagram is invariant under the permutations $2\leftrightarrow4$ and $3\leftrightarrow5$. Although this is not obvious from the Grassmannian integral formula, it will be manifest in the spinorial expression. The Grassmannian integral formula for this diagram is

\begin{equation}\mathcal{D}^{(\tilde{4}+\tilde{4})}_{6,3} = \res_{(356) = 0}\int d^{3\times 6}\Omega_7\frac{[24]\br{24}[35]\br{35}(123)(456)}{(146)(245)(236)(124)(356)}.
\end{equation}
To evaluate the Grassmannian integral, choose
\begin{equation}
C = \begin{pmatrix}
\lambda_1	&\lambda_2	&\lambda_3	&\lambda_4	&\lambda_5	&\lambda_6\\
[24]	&[41]	&0	&[12]	&0	&0
\end{pmatrix},
\end{equation}
which gives
\begin{eqnarray}
\overline{\mathcal{D}}^{(\tilde{4}+\tilde{4})}_{6,3} &\equiv& \overline{\mathcal{D}}^{(124)}_{6,3}, \nonumber \\
&=& \frac{[35]\br{24}\,\delta^{(0|7)}\left([24]\eta_1 + [41]\eta_2 + [12]\eta_4\right)}{s_{124}[12][14]\br{56}\br{36}[1|2+4|3\rangle[1|2+4|5\rangle[4|1+2|6\rangle[2|1+4|6\rangle}.
\label{eqn:124Term}
\end{eqnarray}
From this spinor expression, it is clear that this term is invariant under the permutations $2\leftrightarrow4$ and $3\leftrightarrow5$.\\

The full amplitude is given by
\eq{
\overline{\M}_{6,3}(1^-,2^+,3^-,4^+,5^-,6^+) = \sum_{\genfrac{}{}{0pt}{2}{2\leftrightarrow4}{3\leftrightarrow5}}\left(\overline{\mathcal{D}}^{3+5}_{6,3} + \overline{\mathcal{D}}^{4+4}_{6,3} + \overline{\mathcal{D}}^{5+3}_{6,3}\right) + \overline{\mathcal{D}}^{\tilde{4} + \tilde{4}}_{6,3}.
}
Combining terms with common $s_{ijk}$ poles gives a sum over nine terms, where we denote the term with a pole in $s_{ijk}$ as $\overline{\mathcal{D}}^{(ijk)}_{6,3}$.
\eqs{
\sum_{\genfrac{}{}{0pt}{2}{2\leftrightarrow4}{3\leftrightarrow5}}\overline{\mathcal{D}}^{3+5}_{6,3}&= \overline{\mathcal{D}}_{6,3}^{(345)}+\overline{\mathcal{D}}_{6,3}^{(325)},\\
\sum_{\genfrac{}{}{0pt}{2}{2\leftrightarrow4}{3\leftrightarrow5}}\overline{\mathcal{D}}^{4+4}_{6,3}&=\overline{\mathcal{D}}_{6,3}^{(123)}+\overline{\mathcal{D}}_{6,3}^{(143)}+\overline{\mathcal{D}}_{6,3}^{(125)}+\overline{\mathcal{D}}_{6,3}^{(145)},\\
\sum_{\genfrac{}{}{0pt}{2}{2\leftrightarrow4}{3\leftrightarrow5}}\overline{\mathcal{D}}^{5+3}_{6,3}&= \overline{\mathcal{D}}_{6,3}^{(561)}+\overline{\mathcal{D}}_{6,3}^{(361)},\\
\overline{\mathcal{D}}^{\tilde{4} + \tilde{4}}_{6,3} &= \overline{\mathcal{D}}_{6,3}^{(124)}.
}

\subsection{Local Coordinates} \label{localcoords}

We will now rewrite the expressions obtained in the previous subsection in terms of momentum twistors. For concreteness, let us first consider the formula for the twisted on-shell diagram in \eqref{eqn:124Term}. Using \eqref{square2}, \eqref{etatochi}, and \eqref{plucker}, the fermionic delta function can be written as follows:
\begin{equation}
\delta^{(0|7)}\left([12]\eta_4 + [41]\eta_2 + [24]\eta_1\right) = \frac{\delta^{(0|7)}\big(\br{34}(\br{6123}\chi_5 + \mathrm{cyc.}) + \br{53}(\br{6123}\chi_4 + \mathrm{cyc.})\big)}{\left(\br{61}\br{12}\br{23}\br{34}\br{45}\right)^7}.
\label{eqn:nonplanarferm}
\end{equation}
Compared to the R-invariants which appear in non-MHV amplitudes of $\mathcal{N}=4$ SYM, this expression is complicated and difficult to interpret geometrically. A more illuminating form can be obtained by defining momentum twistors with respect to permuted momenta. Note that we are not actually permuting momenta of the amplitude, so this should be thought of as a passive transformation as we will spell out below. In general, this transformation acts like a permutation of momentum labels, a linear transformation of region momenta, and a non-linear transformation of twistors:
\begin{align}
\left.\ensuremath{p_{i}}\right|_{\mathcal{P}}&=p_{\mathcal{P}(i)}, \nonumber \\
\left.\left(x_{i}-x_{i+1}\right)\right|_{\mathcal{P}}&=\left.p_{i}\right|_{\mathcal{P}}=p_{\mathcal{P}(i)}, \nonumber \\
\left.Z_{i}\right|_{\mathcal{P}}=\left.\left(\lambda_{i},x_{i}\cdot\lambda_{i}\right)\right|_{\mathcal{P}}&=\left(\lambda_{\mathcal{P}(i)},\left.x_{i}\right|_{\mathcal{P}}\cdot\lambda_{\mathcal{P}(i)}\right),
\label{permap}
\end{align}
where the first line is used to express the right hand side of the other two lines. For concreteness, let $\mathcal{P}$ act on momentum labels as follows: 
\begin{equation}
\mathcal{P}=\left(\begin{array}{cccccc}
1 & 2 & 3 & 4 & 5 & 6\\
1 & 2 & 5 & 6 & 3 & 4
\end{array}\right).
\end{equation}
Using \eqref{permap}, the fermionic delta function in \eqref{eqn:nonplanarferm} takes a much simpler form
\eqs{
\delta^{(0|7)}\left([12]\eta_{4}+[41]\eta_{2}+[24]\eta_{1}\right)
&=\left.\delta^{(0|7)}\left([61]\eta_{2}+[12]\eta_{6}+[26]\eta_{1}\right)\right|_{\mathcal{P}},\\
&=\left.\frac{\delta^{(0|7)}(\br{5612}\chi_{3}+\mathrm{cyc.})}{(\br{56}\br{61}\br{12}\br{23})^{7}}\right|_{\mathcal{P}},
}
which can be written in terms of an R-invariant defined in \eqref{Rinvariant}.

Now let's consider a spurious pole in appearing in \eqref{eqn:124Term}:
\begin{equation}
[1|2+4|5\rangle = -\frac{\br{53}(\br{24}\br{6123} + \br{32}\br{6124}) + \br{23}\br{34}\br{5612}}{\br{61}\br{12}\br{23}\br{34}}.
\end{equation} 
In terms of the momentum twistors defined with respect to the permutation $\mathcal{P}$ above, this also takes a much more compact and geometrical form:
\eq{
[1|2+4|5\rangle = [1|2+6|3\rangle \big|_{\mathcal{P}}= \frac{\br{36}\br{5612} + \br{56}\br{6123}}{\br{56}\br{61}\br{12}} \big|_{\mathcal{P}} = \frac{\langle 612|I|356\rangle}{\br{56}\br{61}\br{12}} \big|_{\mathcal{P}},
}
Remarkably, this type of simplification occurs for all of the spurious poles we encounter in the 6-point NMHV amplitude, which makes their cancellation much more transparent, as we show in the next subsection. In Yang-Mills theory, such cancellations were explained by interpreting the amplitude as the volume of a polytope in momentum twistor space. We therefore expect a similar geometric picture for gravity, although it will be more complicated for reasons we will now explain.

The main complication is that we cannot simultaneously simplify all on-shell diagrams using a global choice of coordinates in momentum twistor space. In order to make progress, we will use different momentum twistor coordinates for different on-shell diagrams, analogous to assigning local coordinates on a manifold. In the present context, local coordinates refer to momentum twistors defined with respect to a certain permutation of external momenta which we will refer to as a {\it{chart}}, and the set of on-shell diagrams described by a given chart will be referred to as a {\it{patch}} \footnote{To make the analogy to a manifold more precise, we should distinguish between the amplitude and the underlying geometric object that it describes. Hence, a patch should really be thought of as a region of the underlying geometric object on which a set of on-shell diagrams is defined.}. The transition functions between different patches are complicated in general, but can deduced by decomposing the respective permutations into adjacent transpositions, as we explain in Appendix \ref{transitionfunctions}. Finally, will refer to the set of all charts we use to describe a scattering amplitude as an {\it{atlas}}. 

\begin{figure}[h]
\centering
\includegraphics[width=\textwidth]{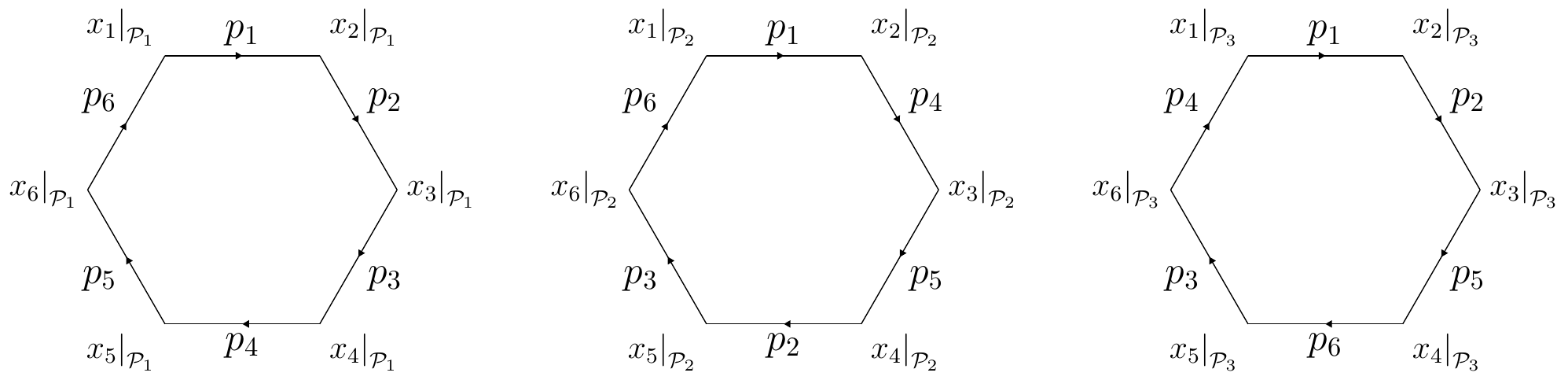}
\caption{Three sets of region momentum variables for the atlas in Table \ref{fig:AtlasChoice}.}
\label{fig:PatchDualVars}
\end{figure}

\begin{figure}[h]
\centering
\includegraphics[width=5cm]{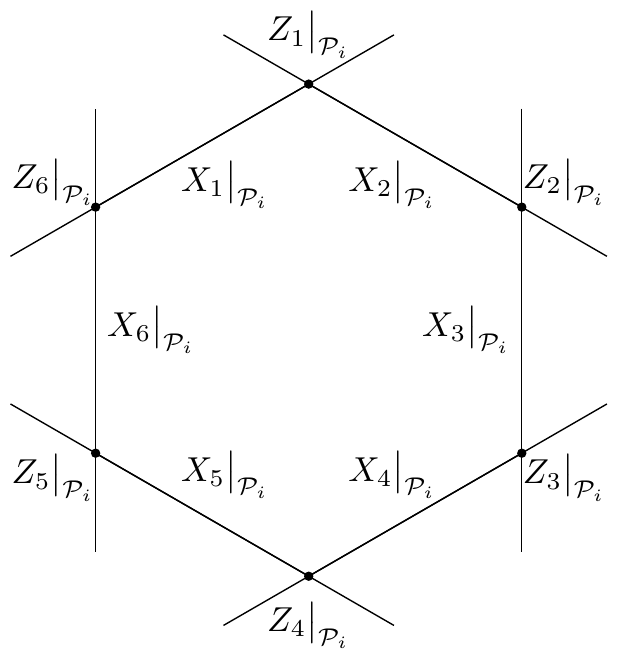}
\caption{Definition of momentum twistors from dual variables on a given patch $\mathcal{P}_i$.}
\label{fig:PatchTwistors}
\end{figure}

The set of all possible charts at $n=6$ is the permutation group $S_6$, but we can quotient by cyclic permutations and the $Z_2$ transformation $\begin{psmallmatrix}1&2&3&4&5&6\\6&5&4&3&2&1\end{psmallmatrix}$, since these permute momentum twistors in a trivial way, leaving a total of 60 charts. Using a Python script, we found that at least three charts are needed to describe the 6-point NMHV amplitude in such a way that all fermionic delta functions can be described by R-invariants. This boils down to looking for the smallest atlases whose charts contain all nine 3-tuples $(i,j,k)$ corresponding to the $s_{ijk}$ poles in the amplitude\footnote{Strictly speaking, the chart itself is the inverse of the permutation we need to apply.}. There are 32 such atlases. Further insisting that one chart corresponds to the identity permutation reduces this to four. We will choose one of these four atlases to demonstrate the relative compactness of the amplitude (the other three atlases lead to similar results). The charts it contains and the on-shell diagrams in each patch are summarised in Table \ref{fig:AtlasChoice}. 
\begin{table}[]
\centering
\begin{tabular}{c|l|l|l}
Patch	&Chart	&Patch Content	&Equations\\
1	&123456	&561, 345, 123	&\req{eqn:561Term}, \req{eqn:345Term}, \req{eqn:123Term}\\
2	&145236	&361, 325, 145	&\req{eqn:561Term}, \req{eqn:345Term}, \req{eqn:123Term}\\
3	&125634	&124, 134, 125	&\req{eqn:124Term}, \req{eqn:123Term}, \req{eqn:123Term}
\end{tabular}
\caption{An atlas for the 6-point NMHV amplitude in momentum twistor space. The permutation associated with each chart is listed in the second column. The on-shell diagrams in each patch are listed in the third column, and are labelled by three momenta appearing the corresponding factorisation channel. The final column lists where the expressions in the third column can be found (some after applying either $3\leftrightarrow 5$ or $2\leftrightarrow 4$ or both).}
\label{fig:AtlasChoice}
\end{table}
We will denote the three charts by the permutations $\mathcal{P}_i$, $i \in\left\{ 1,2,3\right\} $. The region momentum coordinates in each patch are depicted in Figure \ref{fig:PatchDualVars}, and the momentum twistors in a generic patch are depicted in Figure \ref{fig:PatchTwistors}.

To start, let us look at the expressions obtained for the first patch in Table \ref{fig:AtlasChoice}. In this case, the momentum twistors are defined with respect to unpermuted momenta and the three contributions from this patch are given by  
\begin{align}
\overline{\mathcal{D}}^{(561)}_{6,3} &= \frac{R^{(7)}_{251}}{\left[\mathrm{PT}(6)\right]^2}\frac{\br{456|I|124}\br{6123}\br{2345} + \br{612|I|345}\br{1234}\br{4562}}{\br{24}\br{612|I|456}\br{456|I|124}\br{245|I|612}\br{612|45|I|3}}  ,\label{eqn:6ptNMHVTwistor_561}\\
\overline{\mathcal{D}}^{(345)}_{6,3} &= \frac{R^{(7)}_{246}}{\left[\mathrm{PT}(6)\right]^2}\frac{\br{6123}\left(\br{234|I|256}\br{3456} + \br{36}\br{2345}\br{2456}\right)}{\br{26}\br{234|I|456}\br{345|I|236}\br{456|I|236}\br{234|I|256}}  ,\label{eqn:6ptNMHVTwistor_345} \\
\overline{\mathcal{D}}^{(123)}_{6,3}  &= \frac{R^{(7)}_{214}}{\left[\mathrm{PT}(6)\right]^2}\frac{\br{3456}}{\br{34}\br{46}\br{234|I|612}\br{346|I|612}} \label{eqn:6ptNMHVTwistor_123}.
\end{align}
We can see that each term contains an R-invariant along with a squared Park-Taylor factor in the denominator, just as we found for the 5-point NMHV amplitude in \eqref{eqn:5ptNMHVwHelicity}. Furthermore, the remaining terms are written in terms of 4-brackets and 6-brackets defined in \eqref{4bracket}, \eqref{6bracket1} and \eqref{6bracket2}. Note that  4-bracket spurious poles are of the form $\br{i\,j{-}1\,j\,j{+}1}$ for $i{+}1\neq j{-}1$ and $i{-}1 \neq j{+}1$, while 6-bracket spurious poles are of the form $\ang{a|I|bc|def}$ or $\ang{abc|I|def}$ where ($a,b,c$) or $(d,e,f)$ are non-adjacent. In the next subsection we will see that the structure of the terms dressing the R-invariants is required by spurious pole cancellation. It would be interesting to have a more systematic understanding of their structure in terms of some underlying geometric object.  

The second patch in Table \ref{fig:AtlasChoice} does not require any additional work. The twistor expressions are identical to those in the first patch using twistors defined with respect to a different ordering:
\begin{align}
\overline{\mathcal{D}}^{(361)}_{6,3} & = \overline{\mathcal{D}}^{(561)}_{6,3}\big|_{\mathcal{P}_2},\\
\overline{\mathcal{D}}^{(325)}_{6,3} & = \overline{\mathcal{D}}^{(345)}_{6,3}\big|_{\mathcal{P}_2},\\
\overline{\mathcal{D}}^{(145)}_{6,3}& = \overline{\mathcal{D}}^{(123)}_{6,3}\big|_{\mathcal{P}_2}.
\end{align}
Finally, the third patch in Table \ref{fig:AtlasChoice} has the following momentum twistor expressions:
\begin{align}
\overline{\mathcal{D}}^{(124)}_{6,3} &= \frac{R^{(7)}_{163}}{\left[\mathrm{PT}(6)\right]^2} \frac{\br{26}\br{456|I|234}\br{5123}\br{5613}}{\br{23}\br{56}\br{612|I|325}\br{612|I|356}\br{4|I|23|561}\br{4|I|65|123}}\big|_{\mathcal{P}_3},\label{eqn:6ptNMHVTwistor_124}\\
\overline{\mathcal{D}}^{(134)}_{6,3} &= \frac{R^{(7)}_{251}}{\left[\mathrm{PT}(6)\right]^2} \frac{\br{1234}\br{4562}}{\br{24}\br{612|I|456}\br{612|I|452}\br{456|I|214}}\big|_{\mathcal{P}_3},\label{eqn:6ptNMHVTwistor_134}\\
\overline{\mathcal{D}}^{(125)}_{6,3} &= \frac{R^{(7)}_{214}}{\left[\mathrm{PT}(6)\right]^2} \frac{\br{4561}\br{6234}}{\br{46}\br{234|I|612}\br{234|I|614}\br{612|I|436}}\big|_{\mathcal{P}_3}\label{eqn:6ptNMHVTwistor_125}.
\end{align}

In summary, it seems natural to describe sugra amplitudes using momentum twistor coordinates which are defined with respect to different permutations of external momenta. This leads to compact expressions in terms of R-invariants dressed with rational functions of 4-brackets and 6-brackets which have geometric interpretations in terms of intersections of lines and planes in momentum twistor space. As we will demonstrate in the next subsection, this point of view will also make the cancellation of spurious poles in the 6-point NMHV amplitude much more transparent, suggesting a geometric interpretation analogous to the polytope picture discovered for Yang-Mills in \cite{Hodges:2009hk}. Given that each pole of the 6-point NMHV amplitude can be written as either a 4-bracket or a 6-bracket, one may ask if this property holds for all possible poles one can write down at six points. We developed an algorithm to answer this question, which finds the simplest possible form for all poles at six points. The algorithm is explained in appendix~\ref{sec:alg}, and implemented in the attached {\sc Mathematica} file. Using the algorithm, we find that some poles must be written as a sum of terms, and cannot be written as a single 4-bracket or 6-bracket. Hence it is non-trivial that the pole structure of the 6-point NMHV amplitude can be written in a simple geometrical way, and we take this as further evidence that there is an underlying geometric object which encodes gravity amplitudes.

\subsection{Spurious Pole Cancellation}

\begin{figure}[h]
\centering
\includegraphics[width=13.5cm]{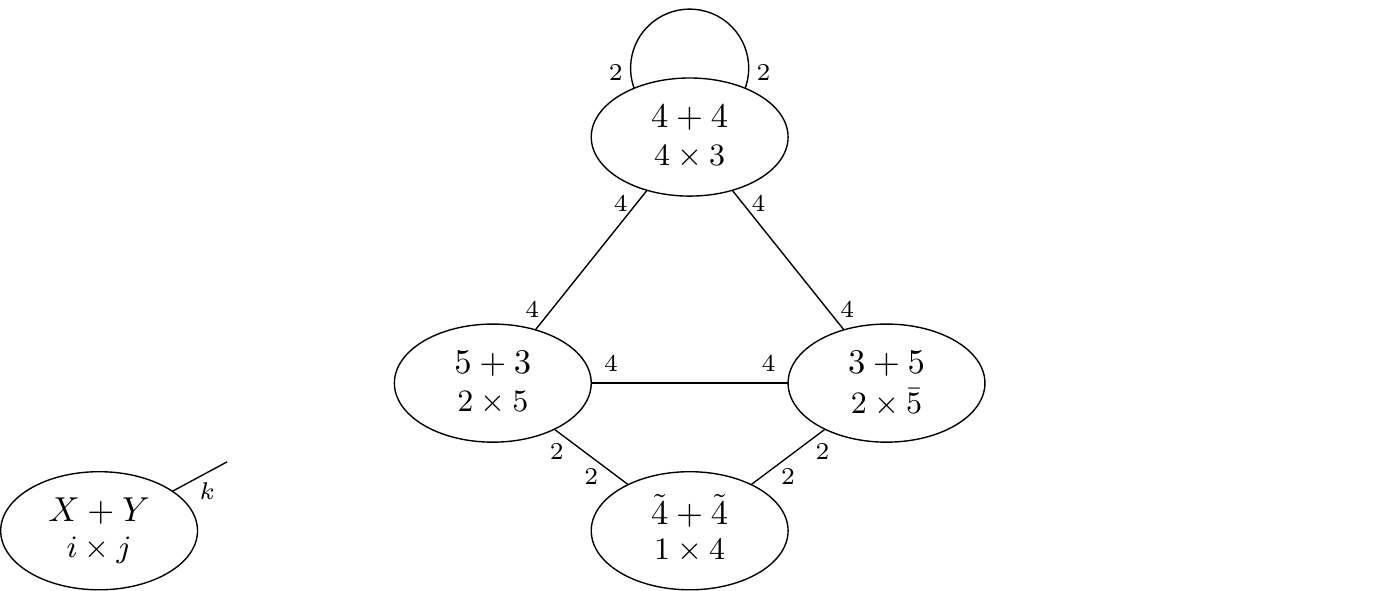}
\caption{Spurious pole structure of the $\N=7$ 6-point NMHV amplitude. The legend on the lower left represents an $X+Y$ diagram topology containing $i$ terms in the permutation sum, each with $j$ spurious poles. The edge represents $k$ pairs of spurious poles which cancel against spurious poles appearing on the other end of the edge.}
\label{fig:Sppoles}
\end{figure}

As first noted in \cite{Hodges:2011wm}, there are 18 different spurious poles in the 6-point NMHV amplitude. Each spurious pole appears twice, giving a total of 36 occurrences. Each of the four diagram topologies has a different set of spurious poles associated with it. These cancel in pairs as shown in Figure \ref{fig:Sppoles}. In principle, there are 18 cancellations to check, however the permutations used to simplify the twistor expressions can also be used to relate all the cancellations which appear on the same edge in Figure \ref{fig:Sppoles}, leaving only six cases to check. We label a given cancellation using the notation $(j|j')$, where the cancellation takes place between two bubbles in Figure \ref{fig:Sppoles} labelled by $i \times j$ and $i' \times j'$. The integers $i$ and $i'$ indicate the number of times the $j$-type and $j'$-type poles occur, respectively. Note that $j$ and $j’$ are in 1-to-1 correspondence with the on-shell diagram topologies denoted $X+Y$ in Figure \ref{fig:Sppoles}, so these two types of labels can be used interchageably when referring to spurious poles. The six cancellations can therefore be labelled $(5|\bar{5}), (5|3), (5|4), (3|3), (\bar{5}|3)$ and $(\bar{5}|4)$. The $(\bar{5}|3)$ and $(\bar{5}|4)$ cancellations can additionally be related to  $(5|3)$ and $(5|4)$ by parity, so there are actually only four cases to check.

We have checked the four cases $(5|\bar{5}), (5|3), (5|4), (3|3)$ analytically using the local coordinates defined in the previous subsection, and in each case the calculation is manifestly supersymmetric and reduces to an application of the Schouten identity, as we illustrate for $(5|\bar{5})$ below. This represents major progress, since in previous work the cancellation of spurious poles was only checked numerically for the graviton component of the superamplitude due to the complexity of the spinorial expressions \cite{Hodges:2011wm}. The cancellations we observe are also very reminiscent of those observed for Yang-Mills amplitudes in \cite{Hodges:2009hk}, which ultimately lead to a new geometric interpretation for NMHV amplitudes as volumes of polytopes. This suggests that a similar interpretation may hold for gravitational amplitudes.

We will now demonstrate the $(5|\bar{5})$ spurious cancellation in Figure \ref{fig:Sppoles} using momentum twistors. To start with, consider the $[5|3+4|2\rangle$ pole shared by $\mathcal{D}^{(561)}$ and $\mathcal{D}^{(345)}$ given in \eqref{eqn:6ptNMHVTwistor_561} and \eqref{eqn:6ptNMHVTwistor_345}, or equivalently \eqref{eqn:561Term} and \eqref{eqn:345Term}. Recall that these expressions arise from $5+3$ and $3+5$ on-shell diagrams, so are associated with the corresponding bubbles in Figure \ref{fig:Sppoles}. Using momentum twistors we can rewrite 
\begin{equation}
[5|3+4|2\rangle = -\frac{\br{4562}}{\br{45}\br{56}},
\end{equation}
so it is clear that we are interested in the behaviour as $\langle4562\rangle \to 0$. This can be viewed as the limit where $Z_2$ approaches the plane defined by $\left\{ Z_{4},Z_{5},Z_{6}\right\}$, denoted $(456)$. In this limit, the remaining twistors $Z_1$ and $Z_3$ become proportional. To see this, consider expanding in the basis $\left\{ Z_{4},Z_{5},Z_{6},Z_{*}\right\}$, where $Z_*$ is an independent reference twistor:
\eqs{
Z_1 &= a_1 Z_4 + b_1 Z_5 + c_1 Z_6 + d_1 Z_*,\\
Z_2 &= a_2 Z_4 + b_2 Z_5 + c_2 Z_6 + d_2 Z_*,\\
Z_3 &= a_3 Z_4 + b_3 Z_5 + c_3 Z_6 + d_3 Z_*. \label{eqn:SpuriousPoleLimit}
}
The limit $\langle4562\rangle \to 0$ corresponds to taking $d_2 \to 0$. In this limit, we can always write $d_3\br{ijk1}= d_1\br{ijk3}$ where $\left\{ 1,3\right\} \notin\left\{ i,j,k\right\}$, since we can neglect any contribution from the $(456)$ plane. Computing the residue of $\mathcal{D}^{(561)}$ and $\mathcal{D}^{(345)}$ of the pole $\br{4562}$ then gives
\eqs{
\mathscr{P} &\equiv \res_{\br{4562}\to0}\left(\overline{\mathcal{D}}^{(561)}_{6,3} + \overline{\mathcal{D}}^{(345)}_{6,3}\right),\\
&=\frac{\br{6123}}{\left[\mathrm{PT}(6)\right]^2\br{24}\br{26}^2}\bigg(\frac{\delta^{(0|7)}\left(\br{4561}\chi_2 + \br{5612}\chi_4 + \br{6124}\chi_5 + \br{1245}\chi_6\right)\br{2345}}{\br{5612}\br{4612}\br{4512}^2\br{4561}^2\left(\br{4612}\br{35} + \br{5612}\br{43}\right)}\\
&\qquad+\frac{\delta^{(0|7)}\left(\br{3456}\chi_2 + \br{5623}\chi_4 + \br{6234}\chi_5 + \br{2345}\chi_6\right)}{\br{2345}\br{2346}\br{2356}\br{3456}^2\left(\br{2345}\br{63} + \br{3456}\br{23}\right)}\bigg),\\
&=\frac{\br{6123}}{\left[\mathrm{PT}(6)\right]^2\br{24}\br{26}^2}\bigg(\frac{\delta^{(0|7)}\left(\br{4563}\chi_2 + \br{5632}\chi_4 + \br{6324}\chi_5 + \br{3245}\chi_6\right)\br{2345}}{\br{5632}\br{4632}\br{4532}^2\br{4563}^2\left(\br{4632}\br{35} + \br{5632}\br{43}\right)}\\
&\qquad+\frac{\delta^{(0|7)}\left(\br{3456}\chi_2 + \br{5623}\chi_4 + \br{6234}\chi_5 + \br{2345}\chi_6\right)}{\br{2345}\br{2346}\br{2356}\br{3456}^2\left(\br{2345}\br{63} + \br{3456}\br{23}\right)}\bigg),\\
&=-\frac{\br{6123}\delta^{(0|7)}\left(\br{3456}\chi_2 + \br{5623}\chi_4 + \br{6234}\chi_5 + \br{2345}\chi_6\right)}{\left[\mathrm{PT}(6)\right]^2\br{24}\br{26}^2\br{2345}\br{2346}\br{2356}\br{3456}^2}\\
&\qquad\qquad\qquad\times\bigg(\frac{1}{\left(\br{4632}\br{35} + \br{5632}\br{43}\right)}-\frac{1}{\left(\br{2345}\br{63} + \br{3456}\br{23}\right)}\bigg),}.
Hence we see that
\eq{
\mathscr{P} \sim \br{2345}\br{63} + \br{3456}\br{23} + \br{6234}\br{53}+\br{5623}\br{43} = 0,  \label{eqn:SUSYpoleCancellation}
}
where we used \eqref{plucker} in the last line. Conveniently, all factors of $d_1$ and $d_3$ cancel in the fraction. We can therefore see that there is no singularity as $\br{4562} \to 0$. The rest of the spurious poles cancel in a similar way. In summary, we find that using momentum twistors defined with respect to different orderings of external momenta provides a very simple and systematic way to prove spurious poles cancelleation in the 6-point NMHV amplitude of $\mathcal{N}=7$ sugra.

\section{Conclusion} \label{conclusion}

Motivated by the beautiful geometric description of scattering amplitudes in planar $\mathcal{N}=4$ SYM, we have tried to follow similar steps for sugra amplitudes. In particular, we first developed an on-shell diagram recursion for $\mathcal{N}=7$ sugra which gives rise to formulas for scattering amplitudes in terms of Grassmannian integrals. This is similar to the on-shell diagram formalism for $\mathcal{N}=8$ supergravity developed in \cite{Heslop:2016plj}, but in $\mathcal{N}=7$ there are two supermultiplets so the diagrams have arrows to indicate helicity flow. The $\mathcal{N}=7$ recursion also appears to have fewer terms than $\mathcal{N}=8$ and automatically incorporates bonus relations for MHV amplitudes. The price to pay for having fewer diagrams is that they contain more closed cycles which can become cumbersome to evaluate at high multiplicity, but we develop a technique to evaluate the diagrams without summing over closed cycles by using a non-canonical gauge-fixing of the Grassmannian integrals. 

Next, we translated our results to momentum twistor space, reproducing Hodges' results for MHV amplitudes \cite{Hodges:2011wm} and obtaining new momentum twistor formulas for non-MHV amplitudes. These formulas are manifestly supersymmetric and written in terms of $\mathcal{N}=7$ R-invariants, analogous to the building blocks for non-MHV amplitudes in $\mathcal{N}=4$ SYM. For the 6-point NMHV superamplitude, this required defining momentum twistors with respect to three different permutations of the external momenta, which can be thought of local coordinates in three different patches. This way of defining momentum twistors was designed to give R-invariants in each patch, but an unexpected consequence of this definition is that the spurious poles greatly simplify and their cancellation becomes very simple to demonstrate. This strongly suggests a geometric interpretation for the cancellation of spurious poles analogous to $\mathcal{N}=4$ SYM and is the main result of this paper.   

There are a number of future directions:
\begin{itemize}
\item Perhaps the most urgent task is to identify the underlying geometry responsible for cancellation of spurious poles in sugra amplitudes. In the context of gluonic amplitudes, this cancellation was made manifest by interpreting 6-point NMHV amplitudes as polytopes in momentum twistor space \cite{Hodges:2009hk}. For sugra amplitudes, identifying the underlying geometry is more challenging because we are only able to describe it using local momentum twistor coordinates, which hide the permutation symmetry of the amplitude. Moreover, the Grassmannian integral formulae for sugra amplitudes have a more complicated form than planar $\mathcal{N}=4$ SYM. It would be interesting to adapt recent work on the geometry of differential forms with non-logarithmic singularities \cite{Benincasa:2020uph} to sugra. It would also be interesting to look for geometric structure in higher-point NMHV amplitudes. New $n$-point formulas recently obtained in \cite{npt} may be useful for this purpose.
\item In $\mathcal{N}=4$ SYM, planar amplitudes are dual to null polygonal Wilson loops \cite{Alday:2007hr,Brandhuber:2007yx,Drummond:2007cf}. In particular, R-invariants correspond to propagators connecting edges of the Wilson loop \cite{Mason:2010yk,CaronHuot:2010ek}. Since R-invariants also appear to play a role in supergravity amplitudes, it would be interesting to look for some analogue of the amplitude/Wilson loop duality in sugra. This was previously found to hold at 4-points in \cite{Brandhuber:2008tf}. Our results suggest this should extend to 5-points, but at higher points one may need to consider multiple Wilson loops for non-MHV amplitudes, one associated with each momentum twistor coordinate patch (see Figure \ref{fig:PatchDualVars}). 
\item Another interesting direction would be to extend the methods developed in this paper to loop amplitudes. When loop amplitudes of planar $\mathcal{N}=4$ SYM are represented in momentum twistor space, they can be expressed in terms of chiral pentagon integrals \cite{ArkaniHamed:2010gh}, which were recently proposed to be building blocks for a dual Amplituhedron \cite{Herrmann:2020qlt}. It would be interesting to see if such integrals can be used to describe sugra amplitudes. While it is not yet clear how to compute loop-level sugra amplitudes using on-shell diagrams, they can be used to compute leading singularities, such as those which were recently studied at 2-loops \cite{Bourjaily:2019iqr}.  
\item Finally, it would be interesting to apply the approach we have developed to sugra and conformal sugra with $\mathcal{N}=4$ supersymmetry. Although these theories have less supersymmetry and the latter is not unitary, their scattering amplitudes have interesting properties and have been studied from various points of view such as twistor string theory \cite{Berkovits:2004jj,Adamo:2012nn,Dolan:2008gc,Adamo:2013tja,Farrow:2018yqf} and the double copy \cite{Bern:2019isl,Johansson:2017srf,Johansson:2018ues}. The amount of supersymmetry in these theories should make it possible to write their amplitudes in terms of the same R-invariants that appeared $\mathcal{N}=4$ SYM.
\end{itemize}
In summary, the study of gravitational amplitudes has revealed many surprises and it seems likely that a more fundamental understanding of their structure remains to be found. 

\section*{Acknowledgements}
We thank Andrew Hodges and Jaroslav Trnka for insightful discussions. AL is supported by a Royal Society University Research Fellowship. CA's studentship is also funded by the Royal Society. 

\newpage

\appendix

\section{Fermionic Delta Functions} \label{app:ferm}

In this Appendix, we will prove the following useful formula for converting fermionic delta functions to momentum twistor space:
\begin{equation}
\delta^{(0|\mathcal{N})}([i\,i{+}1]\eta_{i{+}2}+{\rm cylic})=\frac{\delta^{(0|\mathcal{N})}\left(\left\langle i{-}1\,i\,i{+}1\,i{+}2\right\rangle \chi_{i{+}3}+{\rm cylic}\right)}{\left(\left\langle i{-}1i\right\rangle \left\langle ii{+}1\right\rangle \left\langle i{+}1i{+}2\right\rangle \left\langle i{+}2i{+}3\right\rangle \right)^{\mathcal{N}}}.
\label{fermidel}
\end{equation}
This first appeared in the context of $\mathcal{N}=4$ SYM \cite{Mason:2009qx}, but we use the $\mathcal{N}=7$ version in this paper. The first step is to plug \eqref{superline} into \eqref{theta} to obtain
\begin{equation}
\eta_i = \frac{\anbr{i}{i{+}1}\chi_{i{-}1} + \anbr{i{+}1}{i{-}1}\chi_i + \anbr{i{-}1}{i}\chi_{i{+}1}}{\anbr{i{-}1}{i}\anbr{i}{i{+}1}}.
\label{etatochi}
\end{equation}
Next, using \eqref{brsimp1} and \eqref{brsimp2} we find that
\begin{equation}
\left[ii{+}1\right]\eta_{i{+}2}+{\rm cyclic}=\frac{\left(\br{i{-}1ii{+}1i{+}2}\br{i{+}2i{+}3}\eta_{i{+}2}+\br{i{+}1i{+}2i{+}3|I|i{-}1ii{+}1}\eta_{i{+}1}+\br{ii{+}1i{+}2i{+}3}\br{ii{+}1}\eta_{i}\right)}{\br{i{-}1i}\br{ii{+}1}\br{i{+}1i{+}2}\br{i{+}2i{+}3}}.
\end{equation}
If we substitute \eqref{etatochi} and expand the 6-bracket using \eqref{6bracket1} then we get
\eqs{
\left[ii{+}1\right]\eta_{i{+}2}+{\rm cyclic} &= \Bigg[\frac{\br{i{-}1ii{+}1i{+}2}}{\br{i{-}1i}\br{ii{+}1}\br{i{+}1i{+}2}}\frac{\br{i{+}1i{+}2}\chi_{i{+}3}+{\rm cyclic}}{\br{i{+}1i{+}2}\br{i{+}2i{+}3}}\\
&\quad +\frac{\br{i{+}1i{+}3}\br{i{-}1ii{+}1i{+}2}+\br{i{+}2i{+}1}\br{i{-}1ii{+}1i{+}3}}{\br{i{-}1i}\br{ii{+}1}\br{i{+}1i{+}2}\br{i{+}2i{+}3}}\frac{\br{ii{+}1}\chi_{i{+}2}+{\rm cyclic}}{\br{ii{+}1}\br{i{+}1i{+}2}}\\
&\quad + \frac{\br{ii{+}1i{+}2i{+}3}}{\br{ii{+}1}\br{i{+}1i{+}2}\br{i{+}2i{+}3}}\frac{\br{i{-}1i}\chi_{i{+}2}+{\rm cyclic}}{\br{i{-}1i}\br{ii{+}1}}\Bigg].
}
Collecting coefficients of the $\chi_i$ and factoring out an overall $(\br{i{-}1i}\br{ii{+}1}\br{i{+}1i{+}2}\br{i{+}2i{+}3})^{-1}$ then gives
\eqs{
\chi_{i{-}1}: 	&\: \br{ii{+}1i{+}2i{+}3},\\
\chi_i:			 &\: \frac{1}{\br{ii{+}1}}(\br{i{-}1ii{+}1i{+}2}\br{i{+}1i{+}3}+\br{i{+}3i{-}1ii{+}1}\br{i{+}1i{+}2}+\br{ii{+}1i{+}2i{+}3}\br{i{+}1i{-}1})\\
				&\qquad=\br{i{+}1i{+}2i{+}3i{-}1},\\
\chi_{i{+}1}:	&\: \frac{1}{\br{ii{+}1}\br{i{+}1i{+}2}}\big(\br{i{-}1ii{+}1i{+}2}(\br{ii{+}1}\br{i{+}2i{+}3}+\br{i{+}1i{+}3}\br{i{+}2i})\\
				&\qquad\qquad+\br{i{-}1ii{+}1i{+}3}\br{i{+}2i{+}3}\br{i{+}2i}+\br{ii{+}1i{+}2i{+}3}\br{i{+}1i{+}2}\br{i{-}1i}\big)\\
				& \qquad = \frac{1}{\br{ii{+}1}}\big(\br{i{-}1i{+}1i{+}2}\br{i{+}3i}+\br{i{+}3i{-}1ii{+}1}\br{i{+}2i}+\br{ii{+}1i{+}2i{+}3}\br{i{-}1i}\big)\\
				&\qquad=\br{i{+}2i{+}3i{-}1i},\\
\chi_{i{+}2} :	&\: \br{i{+}3i{-}1ii{+}1},\\
\chi_{i+3}: 		&\:  \br{i{-}1ii{+}1i{+}2},\\
}
from which \eqref{fermidel} follows.

\section{Details of 6pt NMHV} \label{6ptdetail}
In this Appendix, we provide additional details about the 6-point NMHV calculation in section \ref{6ptnmhvos}. We continue will denote $\Delta = \J_C^{-1}$ and $\delta=\delta^{(6|21)}\left(C\cdot\tilde{\lambda}\right)\delta^{6}\left(\lambda\cdot C^{\perp}\right)$, where $ \J_C$ is the Jacobian associated with closed cycles defined in section \ref{sec:algorithm}.
\subsection*{3+5}
\begin{figure}[H]
\centering
\includegraphics[width=6.5cm]{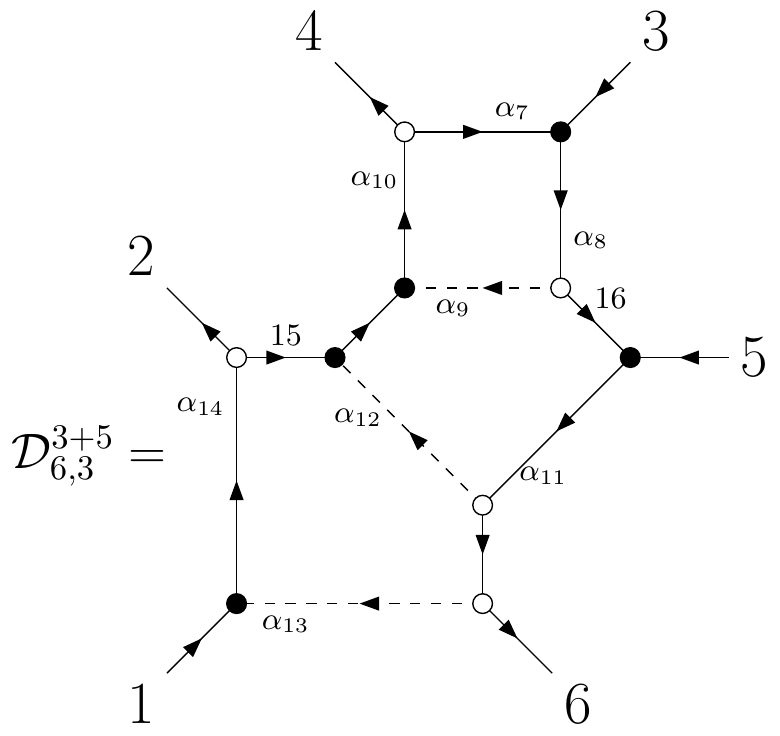}
\caption{`3+5' on-shell diagram to be summed over $2\leftrightarrow4$ and $3\leftrightarrow5$. Edge variables are indicated.}
\label{35osd}
\end{figure}
First we consider the $3+5$ diagram in Figure \ref{35osd}. This diagram gives the C-matrix and Jacobian
\eqs{
C &= \begin{pmatrix}
1	&-\alpha_{14}	&-\alpha_7\alpha_{10}\alpha_{14}	&-\alpha_{10}	\alpha_{14}&0	&0\\
0	&\tilde{\alpha}	&1 - \alpha_7\alpha_8\alpha_9\alpha_{10}	&-\alpha_8\alpha_9\alpha_{10}	&-\alpha_8	&	0\\
-\alpha_{11}\alpha_{13}	&0	&-\alpha_7\alpha_{10}\alpha_{11}\alpha_{12}	&\alpha_{10}\alpha_{11}\alpha_{12}	&1	&-\alpha_{11}
\end{pmatrix},\\
\J_C &= (1 - \alpha_7\alpha_8\alpha_9\alpha_{10} - \alpha_7\alpha_8\alpha_{10}\alpha_{11}\alpha_{12} - \alpha_7\alpha_8\alpha_{10}\alpha_{11}\alpha_{13}\alpha_{14}),
}
where we have inserted an extra $\tilde{\alpha}$ to ensure that $(612)\neq 0$. The brackets from the vertex factors are
\eqs{
\sqbr{2}{15} &= \frac{1}{\alpha_{14}}[21],\\
[47] &= \alpha_7[43],\\
\br{73} &=  \br{43},\\
\anbr{16}{5} &= \frac{1}{\alpha_{11}}\br{65}.
}
As an integral over edge variables, the diagram is then
\eqs{
\mathcal{D}^{3+5}_{6,3} &=\res_{(612)=0} \int \prod_{i = 7}^{14}\frac{d\alpha_i}{\alpha_i^2}\frac{d\tilde{\alpha}}{\tilde{\alpha}}\frac{\delta}{\alpha_8\alpha_{10}\alpha_{11}\alpha_{14}}\frac{\J_C^3\J}{\J}\frac{\alpha_7\alpha_9\alpha_{12}\alpha_{13}}{\alpha_{11}\alpha_{14}}\br{34}[34]\br{56}[12],\\
&= \res_{(612)=0}\int\frac{d^{3\times 6}C}{\gl(3)}\frac{\delta\,\br{34}[34]\br{56}[12]}{\Delta^9 \alpha_7\alpha_8^5\alpha_9\alpha_{10}^6\alpha_{11}^8\alpha_{12}\alpha_{13}\alpha_{14}^7\tilde{\alpha}},\\
&=  \res_{(612)=0}\int\frac{d^{3\times 6}C}{\gl(3)}\frac{\delta\,\br{34}[34]\br{56}[12]}{(124)(234)(345)(356)(561)(612)(456)(346)(256)},\\
&= \res_{(612)=0}\int d\Omega_7^{3\times6}\frac{\br{34}[34]\br{56}[12]}{(256)(346)(356)}\frac{(123)}{(124)}.
\label{35result}
}

\subsection*{5+3}
We can obtain the result from \eqref{35result} using the fact that the 3+5 and 5+3 diagrams are parity conjugates. In particular, we need to exchange square and angle brackets, substitute $(ijk) \to \epsilon_{ijklmn}(lmn)$ and apply the permutation $\mathcal{P}=\begin{pmatrix}1	&2	&4	&3	&5	&6\\6	&5	&3	&4	&2	&1\end{pmatrix}$. This gives
\eqs{
\mathcal{D}^{5+3}_{6,3} &= \res_{(345)\to0}\int \frac{d^{3\times6}C}{\gl(3)}\frac{\delta\,[34]\br{34}[56]\br{12}}{(356)(561)(612)(124)(234)(345)(123)(125)(134)}\bigg|_{\mathcal{P}},\\
&= \res_{(432)\to0}\int\frac{d^{3\times6}C}{\gl(3)}(-1)\frac{\delta\,\br{34}[34]\br{56}[12]}{(421)(216)(165)(653)(543)(432)(654)(652)(643)},\\
&= \res_{(432)\to0}\int d\Omega_7^{3\times6}\frac{\br{34}[34]\br{56}[12]}{(256)(346)(356)}\frac{(123)}{(124)}.
}

\subsection*{4+4}
\begin{figure}[H]
\centering
\includegraphics[width=8cm]{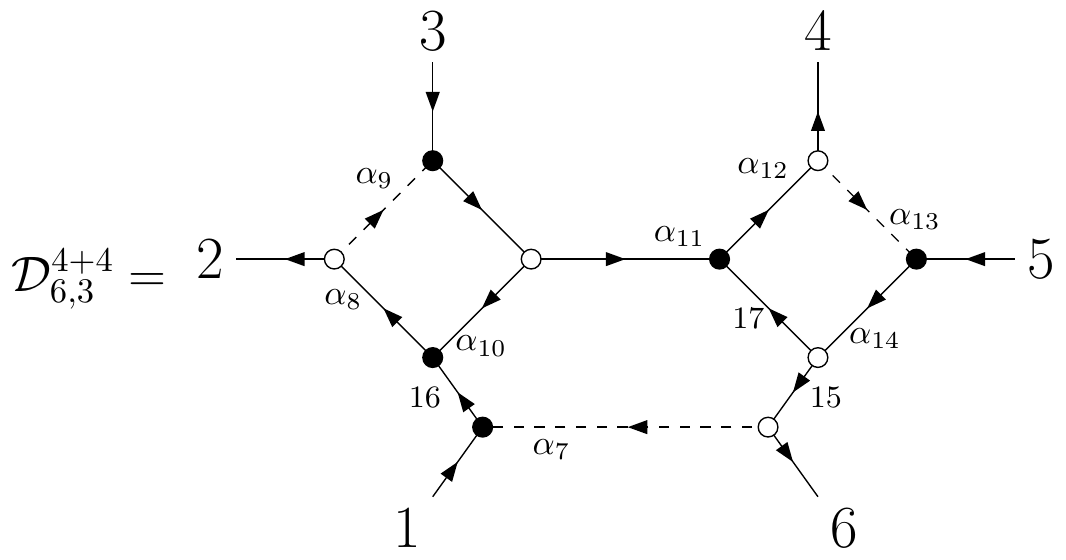}
\caption{`4+4' on-shell diagram to be summed over $2\leftrightarrow4$ and $3\leftrightarrow5$. Edge variables are indicated.}
\label{44osd}
\end{figure}
Next, we consider the $4+4$ diagram in Figure \ref{44osd}. This gives the C-matrix and Jacobian
\eqs{
C &= \begin{pmatrix}
1	&-\alpha_8	&-\alpha_8\alpha_9		&0	&0	&\tilde{\alpha}\\
-\alpha_{10}	&0	&1	&-\alpha_{11}\alpha_{12}	&-\alpha_{11}\alpha_{12}\alpha_{13}	&0\\
-\alpha_7\alpha_{14}	&0	&0	&-\alpha_{12}\alpha_{14}	&1 - \alpha_{12}\alpha_{13}\alpha_{14}	&\alpha_{14}
\end{pmatrix},\\
\J_C &= (1 - \alpha_8\alpha_9\alpha_{10} - \alpha_{12}\alpha_{13}\alpha_{14} - \alpha_7\alpha_8\alpha_9\alpha_{11}\alpha_{12}\alpha_{13}\alpha_{14} + \alpha_8\alpha_9\alpha_{10}\alpha_{12}\alpha_{13}\alpha_{14}).
}
This diagram corresponds to a residue around the pole $(456) = 0$. The bracket factors for the diagram are
\eqs{
\anbr{10}{16} &= \frac{1}{\alpha_8}\anbr{10}{2} = \frac{1}{\alpha_8}\br{32},\\
\sqbr{11}{10} &= \alpha_{10}[31],\\
\anbr{17}{11} &= \anbr{6}{11} = \frac{1}{\alpha_{11}\alpha_{12}}\br{64},\\
\sqbr{15}{17} &= \frac{1}{\alpha_{14}}\sqbr{5}{17} = \frac{\alpha_{12}}{\alpha_{14}}[54].
}
The diagram then evaluates to
\eqs{
\mathcal{D}^{(4+4)}_{6,3} &= \res_{(456)=0}\int\prod_{i=7}^{14}\frac{d\alpha_i}{\alpha_i^2}\frac{d\tilde{\alpha}}{\tilde{\alpha}}\frac{\delta}{\alpha_8\alpha_{12}\alpha_{14}}\frac{\J_C^3\J}{\J}\frac{\alpha_7\alpha_9\alpha_{10}\alpha_{13}}{\alpha_8\alpha_{11}\alpha_{14}}[13][45]\br{23}\br{46},\\
&=  \res_{(456)=0}\int\frac{d^{3\times6}C}{\gl(3)}\frac{\delta\,[13][45]\br{23}\br{46}}{\Delta^9\alpha_7\alpha_8^7\alpha_9\alpha_{10}\alpha_{11}^5	\alpha_{12}^6\alpha_{13}\alpha_{14}^7\tilde{\alpha}},\\
&= \int d^{3\times6}\Omega_7\frac{[13][45]\br{23}\br{46}}{(236)(246)^2}.
}

\subsection*{$\bf{\tilde{4}+\tilde{4}}$}
\begin{figure}[H]
\centering
\includegraphics[width=8cm]{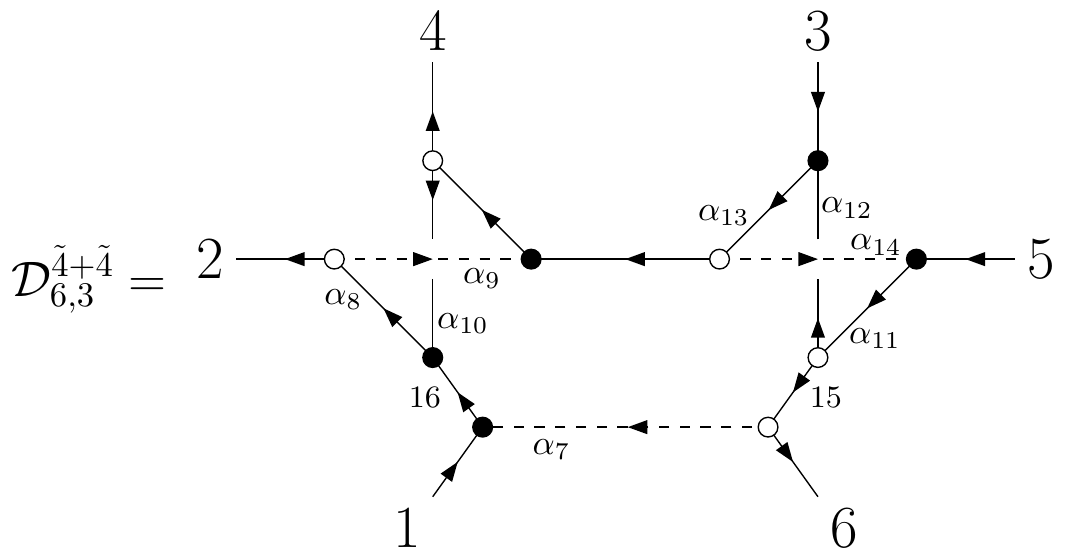}
\caption{`$\tilde{4}$+$\tilde{4}$' on-shell diagram with edge variables indicated.}
\label{4t4tosd}
\end{figure}
Finally, we compute the non-planar $\tilde{4}+\tilde{4}$ diagram in Figure \ref{4t4tosd}. The C-matrix and Jacobian are given by
\eqs{
C &=\begin{pmatrix}
1 - \alpha_8\alpha_9\alpha_{10}	&-\alpha_8	&0	&-\alpha_8\alpha_9	&\tilde{\alpha}\\
-\alpha_{10}\alpha_{13}	&0	&1	-\alpha_{13}	&-\alpha_{13}\alpha_{14}	&0\\
-\alpha_7\alpha_{11}	&0	&-\alpha_{11}\alpha_{12}	&0	&1	&-\alpha_11
\end{pmatrix},\\
\J_C &= (\Delta_1\Delta_2)^{-1} \equiv (1 - \alpha_8\alpha_9\alpha_{10})(1 - \alpha_{11}\alpha_{12}\alpha_{13}\alpha_{14}).
}
Moreover, the bracket factors are
\eqs{
\anbr{10}{16} &= \frac{1}{\alpha_8}\anbr{2}{16} = \frac{1}{\alpha_8}\br{24},\\
\sqbr{10}{4} &=\Delta_1\alpha_8\alpha_{10}[24],\\
\anbr{3}{12} &=\Delta_2\alpha_{11}\br{35},\\
\sqbr{15}{12} &= \alpha_{12}\sqbr{15}{3} = \frac{\alpha_{12}}{\alpha_{11}}[53].
}
Putting it all together, we obtain
\eqs{
\mathcal{D}^{(\tilde{4}+\tilde{4})}_{6,3} &= \res_{(356) = 0} \int\prod_{i=7}^{14}\frac{d\alpha_i}{\alpha_i^2}\frac{d\tilde{\alpha}}{\tilde{\alpha}}\frac{\J_C^3\J}{\J}\frac{\delta}{\alpha_8\alpha_{11}\alpha_{13}}\Delta_1\Delta_2\alpha_7\alpha_9\alpha_{10}\alpha_{12}\alpha_{14}\br{24}[24]\br{35}[53],\\
&= \res_{(356) = 0} \int \frac{d^{3\times6}C}{\gl(3)}\frac{\delta\,\br{24}[24]\br{35}[53]}{\Delta_1^8\Delta_2^7\alpha_7\alpha_8^6\alpha_9\alpha_{10}\alpha_{11}^6\alpha_{12}\alpha_{13}^6\alpha_{14}\tilde{\alpha}},\\
&= \res_{(356) = 0} \int \frac{d^{3\times6}C}{\gl(3)}\frac{\delta\,\br{24}[24]\br{35}[35]}{(124)(234)(345)(356)(561)(612)(146)(236)(245)},\\
&= \mathcal{D}^{(\tilde{4}+\tilde{4})}_{6,3} = \res_{(356) = 0}\int d^{3\times 6}\Omega_7\frac{[24]\br{24}[35]\br{35}(123)(456)}{(146)(245)(236)(124)(356)}.
}

\section{Momentum Twistor Transition Functions} \label{transitionfunctions}
In this Appendix, we will explain how to relate momentum twistors which are defined with respect to different permutations of momenta, which arise in section \ref{localcoords}. First note that cyclic permutations ($i \to i+1$) and reflections ($i \to n-i$) simply permute the momentum twistors in a trivial way. The first non-trivial case is a permutation which exchanges two legs. For concreteness, let us consider the case
\eq{
\mathcal{P} = \begin{pmatrix}
1	&2	&\dots&	n{-}1	&n\\
1	&2	&\dots&	n 			&n{-}1\\
\end{pmatrix},
}
where we have swapped the final two momenta. We will denote the transformed momenta as $p_n\big|_{\mathcal{P}} = p_{n{-}1}$ and vice versa. We can then consider new region momentum and momentum twistor coordinates defined via \eqref{permap}. In particular, this permutation only changes a single $x_i$ so we can write
\begin{equation}
x_i \big|_{\mathcal{P}} = \left\{
\begin{array}{cl}
x_i &, i \neq n,\\
x_{n{-}1} - p_n &, i=n.
\end{array} \right.
\label{eqn:dualvar_perm}
\end{equation}
This means only two twistor variables transform:
\eqs{
Z_i \big|_{\mathcal{P}} &= Z_i ,\qquad i < n{-}1\\
Z_{n{-}1} \big|_{\mathcal{P}}&= \begin{pmatrix}
\lambda_n\\
\lambda_n x_{n{-}1}
\end{pmatrix},\\
Z_{n} \big|_{\mathcal{P}} &= \begin{pmatrix}
\lambda_{n{-}1}\\
\lambda_{n{-}1}(x_{n{-}1} - p_n)
\end{pmatrix},
}
which we can rewrite to put both new twistors on an equal footing:
\eqs{
Z_{n{-}1} \big|_{\mathcal{P}} &= Z_{n} - \anbr{n{-}1}{n}\begin{pmatrix}
0\\
\ltilde{n{-}1}
\end{pmatrix} =  Z_{n} - \anbr{n{-}1}{n}I Z_{n{-}1}\\
Z_n \big|_{\mathcal{P}} &= Z_{n{-}1} - \anbr{n{-}1}{n}\begin{pmatrix}
0\\
\ltilde{n}
\end{pmatrix}  = Z_{n{-}1} - \anbr{n{-}1}{n}I Z_{n},
}
where $I Z_{i}$ is the infinity twistor contracted with $Z_i$.

Now let's consider how twistor brackets transform. Using the rules derived above, we find that
\eqs{
\twbr{a}{b}{c}{n{-}1} \big|_{\mathcal{P}} &= \epsilon_{ABCD}Z_a^AZ_b^BZ_c^C\left.Z_{n{-}1}^{D}\right|_{\mathcal{P}},\\
&= \br{abcn} + \anbr{n{-}1}{n}\left(-\epsilon_{ABCD}Z_a^AZ_b^BZ_c^C\begin{pmatrix}0\\\ltilde{n{-}1}\end{pmatrix}^D\right),\\
&= \br{abcn} + \anbr{n{-}1}{n}\left(-\epsilon_{ABCD}Z_a^AZ_b^BZ_c^C I^{DE} \frac{Z^F_{n{-}2}Z^G_{n{-}1}Z^H_n\epsilon_{EFGH}}{\anbr{n{-}2}{n{-}1}\anbr{n{-}1}{n}}\right),\\
&= \br{abcn} - \frac{1}{\anbr{n{-}2}{n{-}1}}\br{abc|I|n{-}2\,n{-}1\,n}.
}
Using analogous methods, we also find that
\begin{equation}
\twbr{a}{b}{c}{n} \big|_{\mathcal{P}} = \twbr{a}{b}{c}{n{-}1} - \frac{1}{\br{n1}}\br{abc|I|n{-}1\,n\,1},
\end{equation}
and 
\eqs{
\twbr{a}{b}{n{-}1}{n} \big|_{\mathcal{P}} = &-\bigg(\twbr{a}{b}{n{-}1}{n} + \frac{\br{abn|I|n{-}2\,n{-}1\,n}}{\anbr{n{-}2}{n{-}1}} \\&-\frac{\br{ab\,n{-}1|I|n{-}1\,n\,1}}{\br{n1}} + \frac{\br{ab}\anbr{n{-}1}{n}\twbr{n{-}2}{n{-}1}{n}{1}}{\anbr{n{-}2}{n{-}1}\br{n1}}\bigg).
}
The transformation rules for 6-brackets can then be deduced from the above rules using \eqref{6bracket1} and \eqref{6bracket2}. Transition functions for more general permutations can be deduced from repeated composition of   the above rules.

The extension to supertwistors is straightforward. Note that the fermionic components
\eq{
\chi_i = \br{i\,\theta_i} = \br{i\,\theta_{i+1}},
}
transform analogously to the $\tilde{\mu}$ components of the momentum twistors since the fermionic dual variables $\theta_i$ transform analogously to the $x_i$ in \eqref{eqn:dualvar_perm}. We therefore obtain
\begin{equation}
\theta_i \big|_{\mathcal{P}}  = \left\{
\begin{array}{cl}
\theta_i &, i \neq n,\\
\theta_{n{-}1} - q_n &, i=n.
\end{array} \right.
\end{equation}
so that
\begin{equation}
\chi_i \big|_{\mathcal{P}}  = \left\{
\begin{array}{cl}
\chi_i &, i < n-1,\\
\chi_n - \br{n{-}1\,n}\eta_{n{-}1} &, i=n{-}1,\\
\chi_{n{-}1} - \br{n{-}1\,n}\eta_{n} &, i=n.
\end{array} \right.
\end{equation}

\section{Simplifying Momentum Twistor Expressions Algorithmically}
\label{sec:alg}

In this Appendix we develop a new computational method for simplifying expressions in momentum twistor space. These expressions generally have many equivalent forms, related by Schouten identities in the 4-brackets as in equation~(\ref{plucker}), and it is a non-trivial problem to find the most simple form. 

In general the question of simplifying algebraic expressions is complicated, partly as it is not always clear how to define what simple means for a given expression. We consider a subset of expressions in momentum twistor space, and make a definition of what it means for one expression to be more simple than another. Based on this, we outline an algorithm to find the simplest form of a given expression. We provide an explicit realisation of the algorithm in {\sc mathematica}, submitted as an auxilliary file with this publication. Some of the routines in the attached file are based on those from \cite{Farrow:2018cqi}.

We start by defining the form of the functions in momentum twistor space that our algorithm will consider. We then generate and check many possible ansatze to see which are able to match an input spinor or twistor expression. We will work with the set of functions which are polynomial in 4-brackets, and whose coefficients are monomials in angle 2-brackets, with integer powers. All polynomial expressions in 4-brackets in momentum twistor space must be homogeneous, such that each term has the same number of 4-bracket factors. Any equivalent forms of such expressions will also be homogenous polynomials in 4-brackets, with the same degree as in the original expression. Note that 6-brackets can be expressed as a linear function of 4-brackets with angle 2-bracket factors as the coefficients as in equations~(\ref{6bracket1})~and~(\ref{6bracket2}), so we can interchange 6-brackets for 4-brackets and the degree of the polynomial remains the same. We will refer to the input function of the algorithm as $f(\lambda,\mu)$.

The input function which we want to simplify and all ansatz that we check it against must have the following form: 
\eq{
\label{eq:checkansatz}
f(\lambda, \mu) = \sum_{i=0}^N a_i(\lambda) b_i(\lambda, \mu),
}
where $b_i$ are monomials of 4-brackets and 6-brackets (for example, $\ang{1234}$ or $\ang{1247}\ang{123|I|457}$), and $a_i(\lambda)$ have the form 
\eq{
\label{eq:formofai}
a_i(\lambda) = c_i \prod_{1\le x<y\le n}\ang{xy}^{\beta_{ixy}},
}
for $c_i \in \mathbb{Q}$, $\beta_{ixy}\in\mathbb{Z}$, and $N$ an integer determined by the form of $f(\lambda, \mu)$. The input functions we consider may be expressed explicitly in this form, or as functions of spinor brackets which would have this form when converting to momentum twistor space. In this Appendix we consider $\lambda$ and $\mu$ both to be $2\times n$ matrices, and we always suppress the particle and spinor indices.

For such functions in momentum twistor space, we make the definition that simpler expressions have smaller $N$. Hence we ask the question; what is the smallest $N$ such that  $f(\lambda, \mu)$ can be written in the form in equation~(\ref{eq:checkansatz}), for some choice of $a_i(\lambda)$ and $b_i(\lambda, \mu)$? The algorithm for answering this question consists of three different parts. The first is an exhaustive scan over all possible ansatze, and is explained in section~\ref{sec:ScanAnsatz}. For each ansatz, the second part of the algorithm checks numerically to see if the ansatz could match the expression provided, and is explained in section~\ref{sec:checkansatz}. The final part uses a functional reconstruction method to give the complete simplified expression, and is explained in section~\ref{sec:reconstruct}.

\subsection{Scanning over different ansatze}
\label{sec:ScanAnsatz}

We minimise $N$ for a given input function $f(\lambda, \mu)$ by constructing all possible sets of ansatze for $N$, and checking numerically if the input function $f(\lambda,\mu)$ is equal to any of the ansatze. We then increase $N$ until an ansatz which matches $f(\lambda, \mu)$ is found. Note that as our input function $f(\lambda, \mu)$ always has an expression of the form in equation~(\ref{eq:checkansatz}), we always know an upper bound on $N$. We explain this exhaustive search part of the algorithm by way of an example. 

Consider the input function $f(\lambda,\mu) = s_{124}$ at six points. Expanding this expression out naively in terms of angle 2-brackets and 4-brackets using equation~(\ref{square2}) gives an expression of the form given in equation~(\ref{eq:checkansatz}), where $N$ is not minimal. We know this is a polynomial of degree one in 4-brackets, so we can consider ansatze with terms of the form either $\ang{abcd}$, or $\ang{abc|I|def}$. For this example, consider 4-bracket ansatze. There are ${}^6 C_4 = 15$ different 4-brackets; denote the set of these to be $X_{6,4}$.

Start with $N=1$. Construct all different possible ansatze of the form given in equation~(\ref{eq:checkansatz}), where $b_1(\lambda,\mu) \in X_{6,4}$, and $a_1(\lambda)$ is an arbitrary function of $\lambda$. There are 15 such ansatze. For each ansatz of the form $a_1(\lambda) b_1(\lambda,\mu)$, check whether it is equal to $s_{124}$ using the method in section~\ref{sec:checkansatz}. If there are any ansatze which satisfy \eqref{eq:checkansatz}, stop the search and reconstruct $a_1$ using the methods in section~\ref{sec:reconstruct}. If there are no solutions for $N=1$, increase to $N=2$. Now choose all possible combinations of $b_1(\lambda,\mu)$ and $b_2(\lambda,\mu)$ from $X_{6,4}$; there are a total of ${}^{15} C_2 = 105$ such ansatze. Check if any of these $b_1(\lambda,\mu)$ and $b_2(\lambda,\mu)$ give a combination such that 
\eq{
s_{124} = a_1(\lambda) b_1(\lambda,\mu) + a_2(\lambda) b_2(\lambda,\mu),
}
stopping the algorithm and reconstructing $a_1(\lambda)$ and $a_2(\lambda)$ if there are, and continuing to $N=3$ if not. 

The algorithm stops either when a simplified form for $f(\lambda,\mu)$ is found, or we when return the form given in the input function. Note that in general, there may be more than one form for $f(\lambda,\mu)$ with a given value of $N$, and this algorithm will return all such equivalent (equally simple) forms. In the example using $f(\lambda,\mu) = s_{124}$, we could instead have tried to match a 6-bracket ansatz. In that case, we would have used the set $X_{6,6}$ containing all ${}^6 C_3 {}^6 C_2 = 300$ distinct 6-brackets at 6 points.

\subsection{Checking the ansatze}
\label{sec:checkansatz}

We check if each ansatz is equal to the input function $f(\lambda, \mu)$ numerically as follows. First, randomly generate one set of numerical $\lambda\in\mathbb{R}^{2\times n}$. Then generate $2N$ sets of numerical $\mu$, and call them $\mu_i^c$, where $c$ runs over 1 and 2. Note that these indices are unrelated to the standard spinor and particle indices on $\mu$, and each component of $\mu_i^c$ is a numerical matrix in $\mathbb{R}^{2\times n}$. Then evaluate the input function $f(\lambda, \mu)$, and each term $ b_i(\lambda, \mu)$ from the ansatz, on each of the numerical points $(\lambda,\mu_i^c)$, and make the following definitions for the results of those evaluations:
\eq{
f_j^c := f(\lambda, \mu_j^c), \qquad\qquad b_{ij}^c := b_i(\lambda, \mu_j^c), \qquad\qquad a_i := a_i(\lambda).
}
We consider $f_j^1$ and $f_j^2$ as two separate numerical vectors in $\mathbb{R}^N$, and similary $b_{ij}^1$ and $b_{ij}^2$ as two numerical matrices in $\mathbb{R}^{N\times N}$, and $a_i$ are as yet undetermined. $a(\lambda)$ does not depend on $\mu$, so $a_i$ has no $c$ index.

Using these definitions we derive a condition to check whether a given ansatz is false. Starting from equation~(\ref{eq:checkansatz}), we see that
\eq{
f(\lambda, \mu) = \sum_{i=0}^N a_i(\lambda) b_i(\lambda, \mu)\qquad \implies \qquad f_j^c = \sum_{i=0}^N a_i b_{ij}^c
\qquad \implies\qquad  \sum_{i=0}^N \left(b_{ij}^c\right)^{-1} f_j^c = a_j,
}
so long as $\det\left(b_{ij}\right)\ne 0$, which we comment on shortly. Now equate the two terms for $c=1$ and $c=2$ to eliminate $a_i$ to see that
\eq{
\label{eq:numericalcheck}
\sum_{i=0}^N \left(b_{ij}^1\right)^{-1} f_j^1 = \sum_{i=0}^N \left(b_{ij}^2\right)^{-1} f_j^2,
}
is a necessary condition for the ansatz to match the input function. Hence if equation~(\ref{eq:numericalcheck}) is not satisfied for a given ansatz, we know that this ansatz is not correct. To check if the ansatz is true, it is sufficient to check whether $(\ref{eq:numericalcheck})$ holds for a number of different randomly generated momentum twistor points $(\mu,\lambda)$. Treating the expression as a polynomial in the (spinor and particle) components of $\mu$ and $\lambda$, we must check the same number of times as there are terms in this polynomial to prove that the ansatz being checked is equal to the input function. As our algorithm relies on an exhaustive scan over all possible ansatze, efficiency is crucial; we therefore check only once and discard ansatz that return false. We can then run repeated checks on any ansatz that did not return false to see if they equal to the input function. 

There is one final subtlety to address, which is what happens when $\det\left(b_{ij}\right) =0$. There are two cases which can lead to this happening. The first is where two or more rows of $b_{ij}$ are collinear. This can only occur by a random chance where two sets of numerical $\mu_{i}^c$ are the same, and is very unlikely to happen. The other case is where two or more columns of $b_{ij}$ are collinear. This happens when an ansatz is degenerate, for example when a Schouten identity exists between five 4-brackets chosen as the $b_i(\lambda,\mu)$. In this case, the check does not return true or false, but rather returns a message stating that $\det\left(b_{ij}\right) = 0$, and that the ansatz is almost certainly degenerate.

\subsection{Reconstructing the coefficients}
\label{sec:reconstruct}

From the previous steps in our algorithm, we have found at least one ansatz of the form $\sum_{i=1}^N a_i(\lambda) b_i(\lambda,\mu)$ which we know to be equal to the input function $f(\lambda,\mu)$, in terms of undetermined functions $a_i$ of the angle 2-brackets, and we know that these functions will always be of the form given in equation~(\ref{eq:formofai}). In this section, we outline an algorithm for finding each $a_i(\lambda)$ by numerical functional reconstruction. To simplify the notation, we will define a new label $K$ which runs over all ${}^n C_2$ ordered pairs of $x,y$, as in equation~(\ref{eq:formofai}). We will also suppress the $i$ index on $a_i(\lambda)$ and consider only a single function $a(\lambda)$ to be reconstructed. To complete the full reconstruction of $\sum_{i} a_i b_i$ for a given ansatz, the algorithm in this section must be run $N$ times for each value of the $i$ index. 

We then solve the following problem; we are given an $a(\lambda)$ which we can evaluate numerically for different values of $\lambda$, but do not know analytically. Using the streamlined notation, $a(\lambda)$ has the following form 
\eq{
\label{eq:reconstructaj}
a(\lambda) = c \prod_{K = 1}^{{}^n C_2}\ang{K}^{\beta_{K}}.
}
We want to find the ${}^n C_2$ coefficients $\beta_{K}\in\mathbb{Z}$, and the overall constant $c\in\mathbb{Q}$. 

We will take the logarithm of equation~(\ref{eq:reconstructaj}) to produce a linear system to solve for the $\beta_{K}$. This requires that we must have that $\ang{K}>0$ for all $K$. To give this condition on $\lambda$, we then use the projective scaling on each individual momentum twistor to fix $\lambda$ to the form $\left(\begin{smallmatrix}
1&1&...&1\\x_1&x_2&...&x_n
\end{smallmatrix} \right)$. We generate random numbers for the $x_i$ and sort them so that $x_1 < x_2 < ... < x_n$; this ensures that all $\ang{K}>0$. Under this construction, we see that $\lambda$ is a positroid \cite{Bourjaily:2012gy}.

We will solve for ${}^n C_2 + 1$ coefficients $\beta_{K}$ and $c$, so we randomly generate ${}^n C_2 + 1$ sets of numerical positroid $\lambda$; we will label these $\lambda_K$ and $\lambda_e$ ($e$ for extra). We then evaluate $a(\lambda)$ and $\ang{K}$ at each of these numerical values of $\lambda$, and make the following definitions for these numerical quantities:
\eq{
a_{K} := a(\lambda_K)\qquad\qquad
a_{e} := a(\lambda_e)\qquad\qquad
\ang{K}_J := \ang{K}\bigg|_{\lambda = \lambda_J}\qquad\qquad
\ang{K}_e := \ang{K}\bigg|_{\lambda = \lambda_e}.
}
Taking logarithms of equation~(\ref{eq:reconstructaj}) we arrive at the following numerical linear system, which gives ${}^n C_2$ equations for the ${}^n C_2$ undetermined variables $\beta_K$: 
\eq{
 \log\left(\frac{a_{K}}{a_{e}}\right) = \sum_{K= 1}^{{}^n C_2}\beta_{K}\log\left(\frac{\ang{K}_J}{\ang{K}_e}\right).
}
The quantity $\log\left(\frac{\ang{K}_J}{\ang{K}_e}\right)$ is a numerical matrix in $\mathbb{R}^{{}^n C_2\times {}^n C_2}$ with indices $K$ and $J$. We invert this matrix to arrive at
\eq{
\beta_{J} = \sum_{K = 1}^{{}^n C_2}\left(\log\left(\frac{\ang{K}_J}{\ang{K}_e}\right)\right)^{-1}  \log\left(\frac{a_{K}}{a_{e}}\right). 
}
Now that we have the $\beta_{J}$ it is simple to extract $c$, for example using
\eq{
c = a_{e} \prod_{K = 1}^{{}^n C_2}\ang{K}_e^{-\beta_{K}}.
}
Finally we rationalize $\beta_J$ and $c$, using a tolerance of $10^{-8}$. At this step we can check to see if we got an answer that matches the form of the expressions in momentum twistor space that we are working with. We know that generally $\beta_J \in \mathbb{Z}$, and $c$ is normally expected to be 1 or -1, but could also be a small integer, or rational number with a small denominator. If the $\beta_J$ and $c$ returned are not of this form, then we conclude that $a(\lambda)$ was not of the form specified in equation~(\ref{eq:reconstructaj}).

We now have all of the necessary components to fully reconstruct a simplified analytical form for $f(\lambda,\mu)$. We run the algorithm from section~\ref{sec:checkansatz} ${}^n C_2 + 1$ times with numerical positroid $\lambda_K$ and $\lambda_e$ to generate the ${}^n C_2 + 1$ numerical values $a_{iK}$ and $a_{ie}$, for each value of $i$. We then run the algorithm from this subsection $N$ times, one for each value of $i$, generating $c_i$ and $\beta_{iK}$. Finally, we substitute $c_i$ and $\beta_{iK} = \beta_{ixy}$ back into equation~(\ref{eq:formofai}), and then substitute the $a_i(\lambda)$ and $b_i(\lambda, \mu)$ found into equation~(\ref{eq:checkansatz}) to give a fully simplified form of the input function $f(\lambda,\mu)$.

\newpage

\bibliography{sugramp}
\bibliographystyle{JHEP}

\end{document}